\documentclass[aps,pra,twocolumn,superscriptaddress,groupedaddress]{revtex4}

\usepackage{amssymb}
\usepackage{color,graphicx}
\usepackage{amsmath}
\usepackage{amsbsy}
\usepackage{amsthm}
\usepackage{bbm}
\usepackage{bm}
\usepackage{epsfig}
\usepackage{lscape}
\usepackage{float}
\usepackage{subfigure}
\usepackage{placeins}
\usepackage{comment}
\usepackage[english]{babel}

\begin{document}

\title{Ultralow mechanical damping with Meissner-levitated ferromagnetic microparticles.}

\author{A. Vinante}
\email{andrea.mistervin@gmail.com}
\affiliation{School of Physics and Astronomy, University of Southampton, SO17 1BJ, Southampton, UK}
\affiliation{Istituto di Fotonica e Nanotecnologie - CNR and Fondazione Bruno Kessler, I-38123 Povo, Trento, Italy}

\author{P. Falferi}
\affiliation{Istituto di Fotonica e Nanotecnologie - CNR and Fondazione Bruno Kessler, I-38123 Povo, Trento, Italy}

\author{G. Gasbarri}
\author{A. Setter}
\author{C. Timberlake}
\author{H. Ulbricht}
\email{h.ulbricht@soton.ac.uk}
\affiliation{School of Physics and Astronomy, University of Southampton, SO17 1BJ, Southampton, UK}

\date{\today}

\begin{abstract}
Levitated nanoparticles and microparticles are excellent candidates for the realization of extremely isolated mechanical systems, with a huge potential impact in sensing applications and in quantum physics.
Magnetic levitation based on static fields is a particularly interesting approach, due to the unique property of being completely passive and compatible with low temperatures. Here, we show experimentally that micromagnets levitated above type-I superconductors feature very low damping at low frequency and low temperature. In our experiment, we detect 5 out of 6 rigid-body mechanical modes of a levitated ferromagnetic microsphere, using a dc SQUID (Superconducting Quantum Interference Device) with a single pick-up coil. The measured frequencies are in agreement with a finite element simulation based on ideal Meissner effect. For two specific modes we find further substantial agreement with analytical predictions based on the image method. We measure damping times $\tau$ exceeding $10^4$ s and quality factors $Q$ beyond $10^7$, improving by $2-3$ orders of magnitude over previous experiments based on the same principle. 
We investigate the possible residual loss mechanisms besides gas collisions, and argue that much longer damping time can be achieved with further effort and optimization. Our results open the way towards the development of ultrasensitive magnetomechanical sensors with potential applications to magnetometry and gravimetry, as well as to fundamental and quantum physics. 
\end{abstract}

\maketitle

Micromechanical and nanomechanical resonators are widely employed in fundamental and applied physics, due to the fact that they can be interfaced to virtually any kind of system. Broadly speaking, we can identify two main groups of applications. A first one is represented by quantum optomechanics \cite{optoreview,martinis} and tests of quantum physics in the large scale limit \cite{romero2,bouwmeester}. A second one is represented by mechanical sensing of weak signals. Well known examples are force microscopy \cite{afm}, force-detected single spins \cite{rugarsingle}, detection of small masses at molecular level \cite{bachtold}, ultrasensitive accelerometry \cite{pike} and gravimetry \cite{rowan,tang}, magnetometry \cite{budker2} and tests of wavefunction collapse models \cite{vinanteCSL}.


For sensing applications a proper figure of merit to compare different systems is the ratio $T / \tau$ where $T$ is the temperature and $\tau $ is the mechanical damping time. In fact, the ultimate limit for sensing is given by the thermal force noise, with spectral density $S_f=4 k_B T m \Gamma$ where $\Gamma= \omega/ Q=2/\tau$ is the energy dissipation rate, $\omega$ is the angular frequency, $Q$ is the quality factor and $m$ is the mass \cite{tau}. It is apparent that, for a given $Q$ factor, lower thermal noise is achieved at lower frequencies.

To date, clamped mechanical resonators optimized for quantum optomechanics at MHz and GHz have achieved ultrahigh $Q$ factors up to $10^9-10^{10}$ \cite{schliesser,painter}. The factor $T/\tau$ can be as low as $10^{-2}$ K/s for cryogenic micromembranes \cite{schliesser}. Similar values have been achieved by aluminum membranes cooled to millikelvin temperatures \cite{lehnert}. At kHz frequencies, the most typical mechanical sensors are micromachined cantilevers for force microscopy \cite{afm,rugar}, which achieve $T/\tau$ as low as $10^{-4}$ K/s at millikelvin temperatures\cite{vinanteCSL2}. Finally, microelectromechanical systems optimized for seismometry and gravimetry at Hz frequencies achieve $Q \approx 10^4$ in vacuum at room temperature\cite{pike,tang}. This means that they can potentially achieve $T/\tau \approx 10^{-4}$ K/s at cryogenic temperatures.

Levitated microparticles and nanoparticles have recently emerged as a very promising alternative to clamped mechanical resonators. Besides flexibility and tunability of mechanical parameters, a particularly attractive feature of levitated systems is the absence of clamping losses. As a consequence, very low levels of dissipation and decoherence are potentially achievable \cite{chang}. Optical levitation, the most popular and developed technique \cite{ashkin}, has been widely demonstrated for force \cite{fb2, geraci, force1}, torque \cite{tongli,force2} and acceleration sensing \cite{monteiro}. In spite of high flexibility, optical levitation suffers from heating induced by photon-absorption and scattering, which prevents low temperature operation and ultimately leads to excess thermal noise and decoherence \cite{romero1}. The lowest $T/\tau$ factor for optically levitated nanoparticles is approximately $10^{-2}$ K/s, which is obtained in low frequency systems \cite{monteiro}.

This suggests that different levitation methods, such as Paul traps~\cite{paul, pontin, teq, hetet}, or magnetic traps~\cite{durso, cinesi, chris, budker1} could outperform optical levitation in applications requiring the lowest possible noise level. In particular, magnetic levitation appears very promising because of the unique combination of two properties: a completely passive trapping by static magnetic fields, and the possibility of using SQUIDs to detect the motion with ultralow power dissipation. As a consequence it appears as the perfect solution for operation at low temperatures. Magnetic levitation can be implemented using diamagnetic \cite{durso, cinesi} or superconducting particles \cite{romero1} in external static fields, or ferromagnetic particles above superconductors \cite{chris, budker1}. Interestingly, these systems have been proposed not only for ultrasensitive force and inertial sensing \cite{romero3,chris}, but also to test quantum mechanics in currently inaccessible regimes \cite{romero2}, to enable quantum technologies such as quantum magnetomechanics \cite{romero1} and acoustomechanics \cite{romero4} and for ultrasensitive magnetometry \cite{budker2}. In particular, it has been suggested that a levitated micromagnet can overcome the standard quantum limitations to the resolution per unit volume of a magnetometer \cite{budker2,mitchell}. 


In this paper we experimentally investigate magnetic levitation of micromagnets using passive superconducting traps, employing SQUIDs for low power position detection. Specifically, we levitate, trap and detect individual ferromagnetic microparticles in a type-I superconducting trap made of lead, and identify rigid body modes in the frequency range $1-400$ Hz. We perform a mechanical characterization at $T=4.2$ K at pressures down to $10^{-5}$ mbar, measuring quality factors in excess of $10^7$ and damping time in excess of $10^4$ s. These results improve over similar previous experiments by more than two orders of magnitude. The value of $T/ \tau \approx 10^{-4}$ K/s is already at the state of the art for microsensors, and can be further improved by cooling to millikelvin temperature.
We investigate the limiting dissipation mechanisms and discuss a number of possible applications, including force and acceleration sensing, gravimetry, magnetometry as well as fundamental and quantum physics experiments. As a case study, we will investigate the potential of using our levitated micromagnet as a torque magnetometer.

\section{Experimental setup and mode identification}

The trap consists of a cylindrical well inside a $99.95\%$-purity Pb block (Figs. 1a and 1b) with $4$ mm diameter and $4$ mm depth. The Meissner surface currents induced by the magnetic particle, combined with gravity, provide the vertical confinement, while the lateral surface provides the horizontal one. No special loading procedure is implemented. The magnetic particle is individually manipulated at room temperature and placed at the bottom of the trap before cooling down. Upon cooldown, Meissner repulsion alone is sufficient to overcome electrical interactions and levitate the particle. We have checked this behaviour for particle radii down to $\sim 30$~$\mu$m.
\begin{figure}[!ht]
\includegraphics[width=8.6cm]{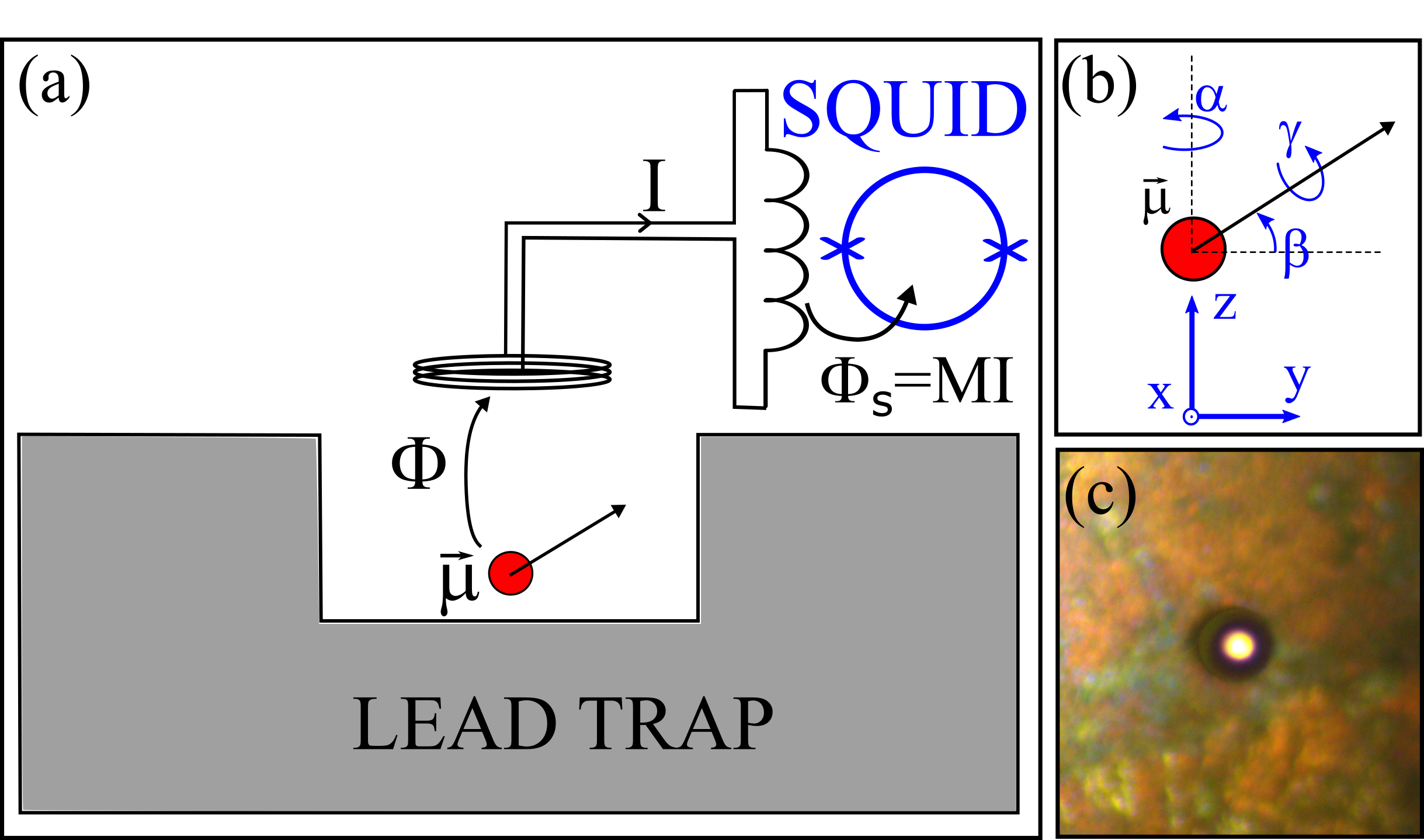}
\caption{(a) Simplified scheme of the trapping and detection technique. A hard ferromagnetic microparticle is levitated inside a cylindrical well in a type-I superconductor, in our case lead. Meissner repulsion combined with gravity keeps the magnetic particle trapped in all directions. Both the translational and the rotational motion of the particle are detected by a superconducting pick-up connected to a dc SQUID. (b) Conventions adopted on the translational and rotational degrees of freedom. Under weak breaking of cylindrical symmetry, all modes are trapped and detected, except $\gamma$. (c) Optical microscope image of a microsphere with radius $R=\left( 27 \pm 1 \right)$ $\mu$m placed on the bottom of the trap. The polycrystalline lead grain boundaries are visible.}  \label{scheme}
\end{figure}

We have experimentally demonstrated stable trapping for a variety of spherical and cylindrical magnetic particles with characteristic size in the range $30-1000$ $\mu$m. Here, we will focus on the smallest particle we have levitated, a microsphere made of a neodymium-based alloy \cite{magnequench} with radius $R=\left(27 \pm 1 \right)$ $\mu$m (Fig.~1c), determined by optical microscope inspection. The microsphere has been fully magnetized in a $10$ T NMR magnet prior to the experiment, with an expected saturated magnetization $\mu_0 M \approx 0.7$ T.

We detect the motion of the particle using a commercial dc SQUID from Quantum Design connected through a single pick-up coil placed above the levitated particle (Fig.~1a). The motion along a given degree of freedom induces a change of the magnetic flux $\Phi$ in the pick-up coil produced by the permanent magnetic dipole $\bm{ \mu}$ of the particle. As the pick-up coil connected to the SQUID input coil constitutes a superconducting loop of total inductance $L$, this results in a current $I$ and a flux $\Phi_S=M I =\left( M/L \right) \Phi$ into the SQUID, with $M$ the mutual inductance between SQUID and input coil. The pick-up coil consists of $6$ loops of NbTi wire, wound around a cylindrical PVC holder with radius $1.5$~mm, coaxial with the trap. The SQUID is enclosed in a separate superconducting shield. The NbTi wires connecting the SQUID input coil and the pick-up coil are pair-twisted and shielded by a lead tube. We can drive the microparticle by means of a second NbTi coil wound on the same pick-up coil holder, connected to an external resistive line. 

Trap and SQUID are enclosed in a indium-sealed copper vacuum chamber which can be dipped in a standard helium transport dewar. A rough mechanical isolation is obtained by hanging the dewar from the laboratory ceiling through a silicone tube. This reduces seismic noise in the bandwidth 1-20 Hz by more than 1 order of magnitude. A cryoperm shield reduces the earth field by a factor of about $10$ at the lead trap location. After evacuating the chamber at room temperature, 1 mbar of helium exchange gas is added for thermalization and pumped out after cooldown. We monitor the pressure in the vacuum chamber with a Pirani-Penning gauge placed at room temperature. Pressure data is corrected for the gas composition by assuming that the gas is pure helium.
\begin{figure}[!ht]
\includegraphics[width=8.6cm]{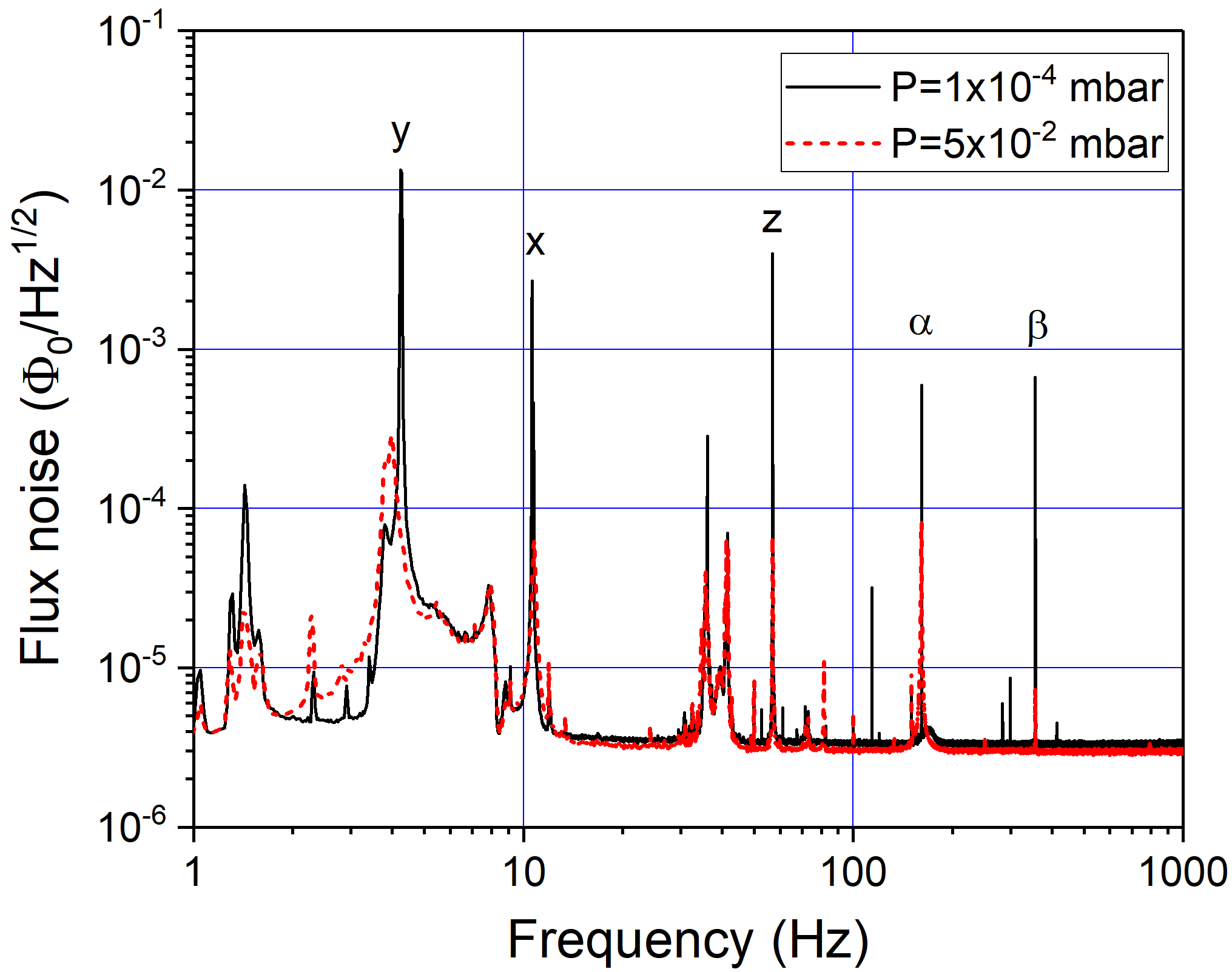}
\caption{Spectrum of the flux measured by the SQUID at two representative nominal pressures, $P=1.0 \times 10^{-4}$~mbar (black continuous line) and $P=5.0 \times 10^{-2}$~mbar (red dotted line). The rigid body mechanical modes of the microsphere are marked by labels.}  \label{spectrum}
\end{figure}

Fig.~2 shows two typical SQUID spectra at low and high pressure. The spectra are averaged over a total time of $4.2 \times 10^4$ s.  Five modes related with the microsphere rigid body motion have been identified, as described below. Other peaks and structures are due to seismic and acoustic noise. We can distinguish the levitated particle modes from spurious peaks thanks to the high quality factor $Q$ and its characteristic dependence on pressure, as discussed below. 

To identify the observed peaks with specific translational and rotational modes of the microparticle, we need a reliable modeling of the system. We parameterize the motion of the particle with the center of mass coordinates $x,y,z$, the librational angle $\beta$ and the azimuthal angle $\alpha$ defined by the permanent magnetic moment $\bm{ \mu}$, and the rotation angle $\gamma$ around $\bm{ \mu}$ (Fig. 1b). The origin is chosen at the center of the bottom surface of the trap. 
For ideal Meissner effect and trap axis aligned with the gravitational field (i.e. with no tilt), the equilibrium position corresponds to $x_0,y_0,\beta_0=0$ and a finite equilibrium height $z_0$ depending on the equilibrium between Meissner repulsion and gravity. There will be no confinement in the $\alpha$ and $\gamma$ angle, so we would expect to detect only $4$ modes. In a real implementation any finite tilt with respect to the Earth gravitational field shifts the horizontal equilibrium position off center, breaking the rotational symmetry around $\alpha$, and shifting all frequencies. A trapped torsional $\alpha$ mode will then show up, leading to $5$ detectable modes. Because of symmetry breaking all $5$ modes are expected to show some finite coupling to the pick-up coil.

In the following, we will mostly focus on two specific modes, namely the $z$ and the $\beta$ ones. Remarkably, the frequency of these modes can be analytically predicted with reasonable accuracy, as a straightforward application of the image method \cite{image1,image2}. This is because the dynamics of these modes depends almost exclusively on the interaction of the microsphere with the bottom surface, as long as the equilibrium height $z_0$ is much smaller than the trap radius. Therefore, the resonance frequency is expected to depend very little on the trap tilt. Indeed, we do observe two modes with relatively stable frequency at $\left( 56.5 \pm 0.1 \right)$~Hz and $\left( 377 \pm 5\right)$~Hz, while the other $3$ modes show frequency variations up to $50 \%$ upon tilting the experimental setup by a few degrees.

According to the image method \cite{image1,image2}, the potential energy of a permanent magnet with magnetic moment $\bm{ \mu}$ and mass $m$, placed above an horizontal infinite superconducting plane, is given by:
\begin{equation}
U=\frac{\mu_0 \mu^2}{64 \pi z^3}\left( 1+\mathrm{sin}^2\beta \right) + mgz
\end{equation}
Upon minimization one finds that the equilibrium position is achieved at $\beta =\beta_0=0$, i.e. with the magnet positioned horizontally, and $z=z_0$, with:
\begin{equation}
z_0=\left( \frac{3 \mu_0 \mu^2}{64 \pi m g }\right)^{\frac{1}{4}}
\end{equation}
The angular resonance frequencies of the $z$ and $\beta$ modes are then easily calculated as $\omega_z=\sqrt{k_z/m}$ and $\omega_{\beta}=\sqrt{k_\beta/I}$, where $I$ is the moment of inertia and $k_z$,$k_\beta$ are the spring constants:
\begin{align}
 & k_z =\!\left[ \frac{d^2 U}{dz^2} \right]_{\left( z, \beta \right)=\left(z_0,\beta_0 \right)}, \\
  &k_\beta =\!\left[ \frac{d^2 U}{d\beta^2} \right]_{\left(z,\beta \right)=\left(z_0,\beta_0 \right) }
\end{align}

Assuming our particle is a homogeneous sphere, the moment of inertia is $I=\frac{2}{5}mR^2$ and we finally obtain: 
\begin{align}
 & \omega_z=\sqrt{\frac {4 g}{z_0}}  \label{wz} \\
 & \omega_\beta=\sqrt{\frac{5 z_0 g}{3 R^2} } \label{wb}
\end{align}
Interestingly, both frequencies can be written as the square root of the ratio of $g$ with an effective length, as in a simple pendulum.
Furthermore, by multiplying Eq.~(\ref{wz}) with Eq.~(\ref{wb}) we derive the following remarkably simple relation:
\begin{equation}
  R=\sqrt{\frac{20}{3}}\frac{g}{\omega_z \omega_\beta}   \label{R}
\end{equation}
which provides the microsphere radius $R$ as a function of the two frequencies, independent of mass density and magnetization.

Plugging the experimental values of the two stable frequencies into Eq.~(\ref{R}), we obtain a radius $R=\left( 30.1 \pm 0.5 \right)$~$\mu$m, which is compatible within $10 \%$ with the radius $R=\left( 27 \pm 1 \right)$~$\mu$m measured by optical inspection. The small discrepancy could be partially explained by a finite tilt, as discussed later, or by deviations from sphericity. Further use of Eqs.~(\ref{wz},\ref{wb}) allows identification of the $z$-mode at $\omega_z/2\pi=\left( 56.5 \pm 0.1 \right)$~Hz and the $\beta$-mode at $\omega_\beta/2\pi=\left( 377 \pm 5 \right)$~Hz. The resonance frequencies are correctly reproduced by using the values $\rho=7430$~kg/m$^3$ for the mass density and $\mu_0 M=0.71$~T for the magnetization, which are consistent with the values provided by the manufacturer. The derived equilibrium height $z_0=311$~$\mu$m is much smaller than the hole radius, which justifies the assumption of infinite plane.

A finite element analysis is required to account for the full dynamics and to predict the frequency of all modes. 
Results from a full 5D finite element simulation of our system based on FEniCS software \cite{fenics} are shown in Appendix 1. The simulation provides the equilibrium position and the resonant frequencies of all 5 modes, as a function of a finite tilt $\theta$ applied along the $x$-axis. The mode frequencies are in reasonable agreement with the experimental results by assuming a tilt of about $3$ degrees. In particular, the $z$ and $\beta$ mode are in substantial agreement with the image method estimation, and show indeed a weak dependence on the tilt, in contrast with the other modes. The low frequencies peaks in the spectrum (Fig.~2) at about $4$ and $10$ Hz are identified as the $y$ and $x$ mode. The remaining peak is identified as the $\alpha$ mode. For the latter, a somewhat larger discrepancy between experiment and simulation is observed, with a predicted value of $\sim 100$ Hz and measured value of about $\sim 150$ Hz. This may indicate some asymmetrical feature in the system which is not accounted for by the model, such as deviations from sphericity or magnetization inhomogeneities.

The simulation also provides an estimation of the equilibrium values of the 5 independent variables. Assuming the tilt along the $x$ axis, the equilibrium configuration is $\left( x,y,z,\beta , \alpha \right) \approx \left( x_0, 0, z_0, 0 , \pi/2 \right)$, with $z_0$ in agreement within $10\%$ with the image method estimation and $x_0$ which is strongly dependent on the tilt.


\section{Characterization of mechanical modes}

We measure the amplitude decay time $\tau$ of a given mode at frequency $f=\omega/2\pi$ by means of standard ringdown measurements. The mode is selectively excited at large amplitude by sending a current signal into the excitation coil, and the ringdown is acquired by a lock-in amplifier with reference tuned on the resonance frequency. $\tau$ is inferred from an exponential fit of the amplitude versus time. The quality factor is calculated as $Q=\pi f \tau$. At the lowest measured pressures $P \sim 10^{-5}$ mbar, $\tau$ is of the order of several hours and the measurement becomes very difficult, due to the interplay between environmental  vibrational noise and the onset of nonlinearities which restrict the available linear dynamic range. Nevertheless, we could perform reliable ringdown measurements of the $\beta$ mode at all pressures, and of the $z$ mode at almost all pressures.
\begin{figure}[!ht]
\includegraphics[width=8.6cm]{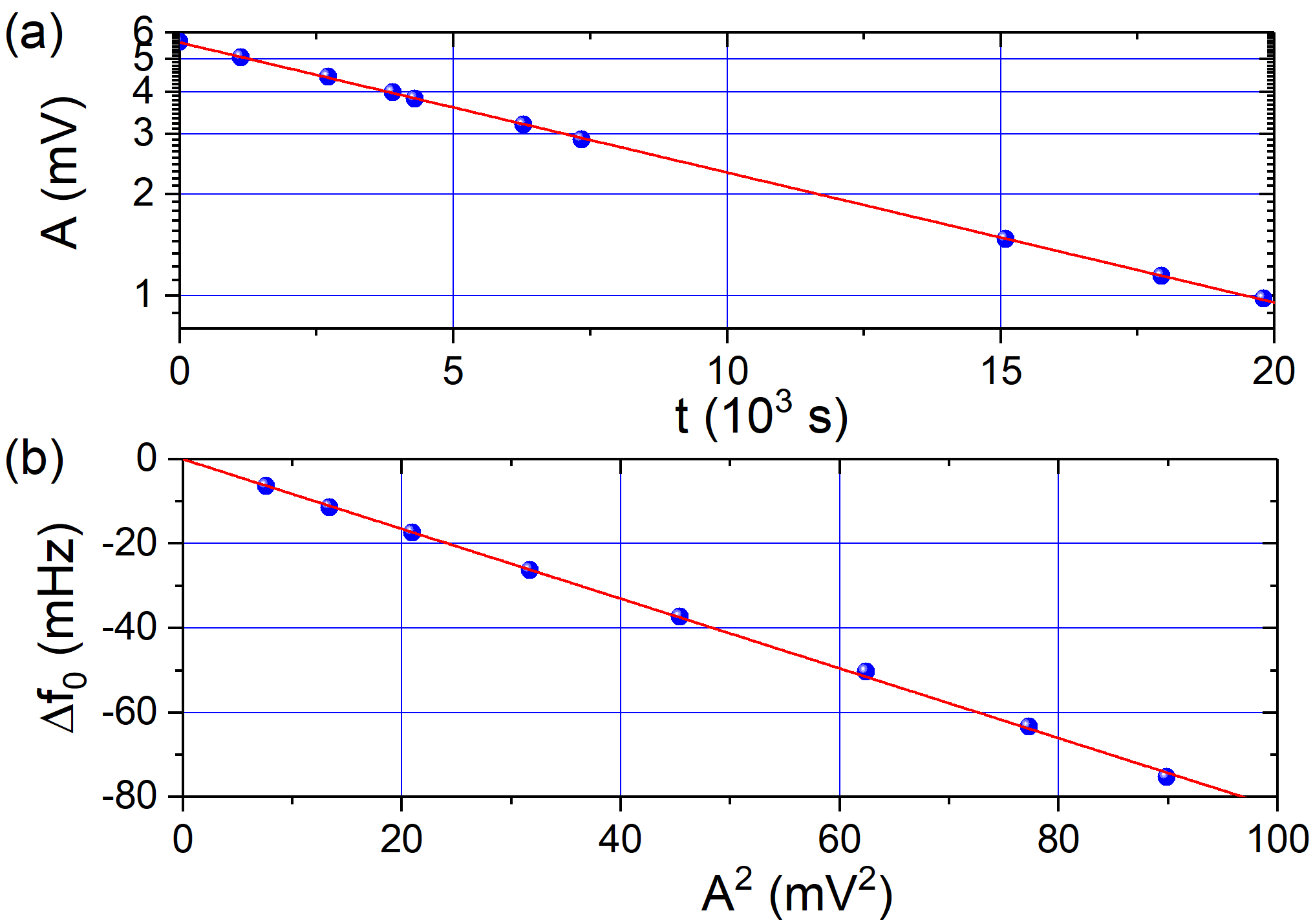}
\caption{(a) Example of ringdown measurement of the $\beta$ mode at the lowest nominal pressure $P=5.7 \times 10^{-5}$ mbar. (b) Frequency shift of the $\beta$ mode as a function of the uncalibrated square amplitude, measured during a ringdown.}  \label{decay1}
\end{figure}

As first example, we show in Fig.~3a a ringdown measurement of the $\beta$ mode at nominal pressure $P=5.7 \times 10^{-5}$~mbar, featuring a decay time $\tau=1.13 \times 10^4$~s and a quality factor $Q=1.34 \times 10^7$.  The ringdown is exponential and does not feature significant damping nonlinearities. Instead, we observe a clear spring softening effect, appearing as a negative frequency shift proportional to the square amplitude, as shown in Fig.~3b. This behaviour is entirely expected, as the trapping potential is intrinsically nonlinear. Based on the analytical potential derived from the image method, we can calculate explicitly the expected frequency shift as a function of the angular amplitude (see Appendix 2 for details). This provides an indicative calibration of the absolute value of the angular motion. The absolute motion for the ringdown of the $\beta$ mode in Fig.~3a corresponds to a range $\beta=3-20 \times 10^{-3}$ rad, while the typical amplitude of the undriven mode after the exponential ringdown corresponds to about $10^{-4}$ rad. This is significantly larger than the mean thermal motion, which can be estimated as $\beta_{th}=\sqrt{k_B T/k_\beta}\approx 6 \times 10^{-6}$ rad, suggesting that the $\beta$ mode is dominated by ambient vibrational noise, for instance seismic noise. 

In Figs.~4a and~4b we show a similar ringdown and frequency shift measurement of the mode $z$. The ringdown is acquired at nominal pressure $P=1.0 \times 10^{-4}$~mbar, and shows a decay time $\tau=1.17 \times 10^4$~s and a quality factor $Q=2.1 \times 10^6$. The frequency shift measurement shows a behaviour similar to the $\beta$ mode, proportional to the square of the oscillation amplitude. Using the expected frequency shift from the image method, we find that the the absolute motion in the ringdown measurement corresponds to a range $z=10-20$ $\mu$m, while the typical amplitude of the undriven mode corresponds to about $0.2$ $\mu$m. Similarly to the $\beta$ mode, the motion of the undriven mode is much larger than the expected mean thermal motion $z_{th}=\sqrt{k_B T/k_z}\approx 0.7$ nm, showing that the $z$ mode is also dominated by ambient vibrations. A further mechanical isolation by at least $3$ orders of magnitude would be needed to suppress excess noise well below the thermal noise. We note the estimation of the absolute amplitude of motion allows us to calibrate the displacement noise of the SQUID detection. For the $z$-mode the noise is of the order of 1 nm/$\sqrt{\mathrm{Hz}}$.
 
\begin{figure}[!ht]
\includegraphics[width=8.6cm]{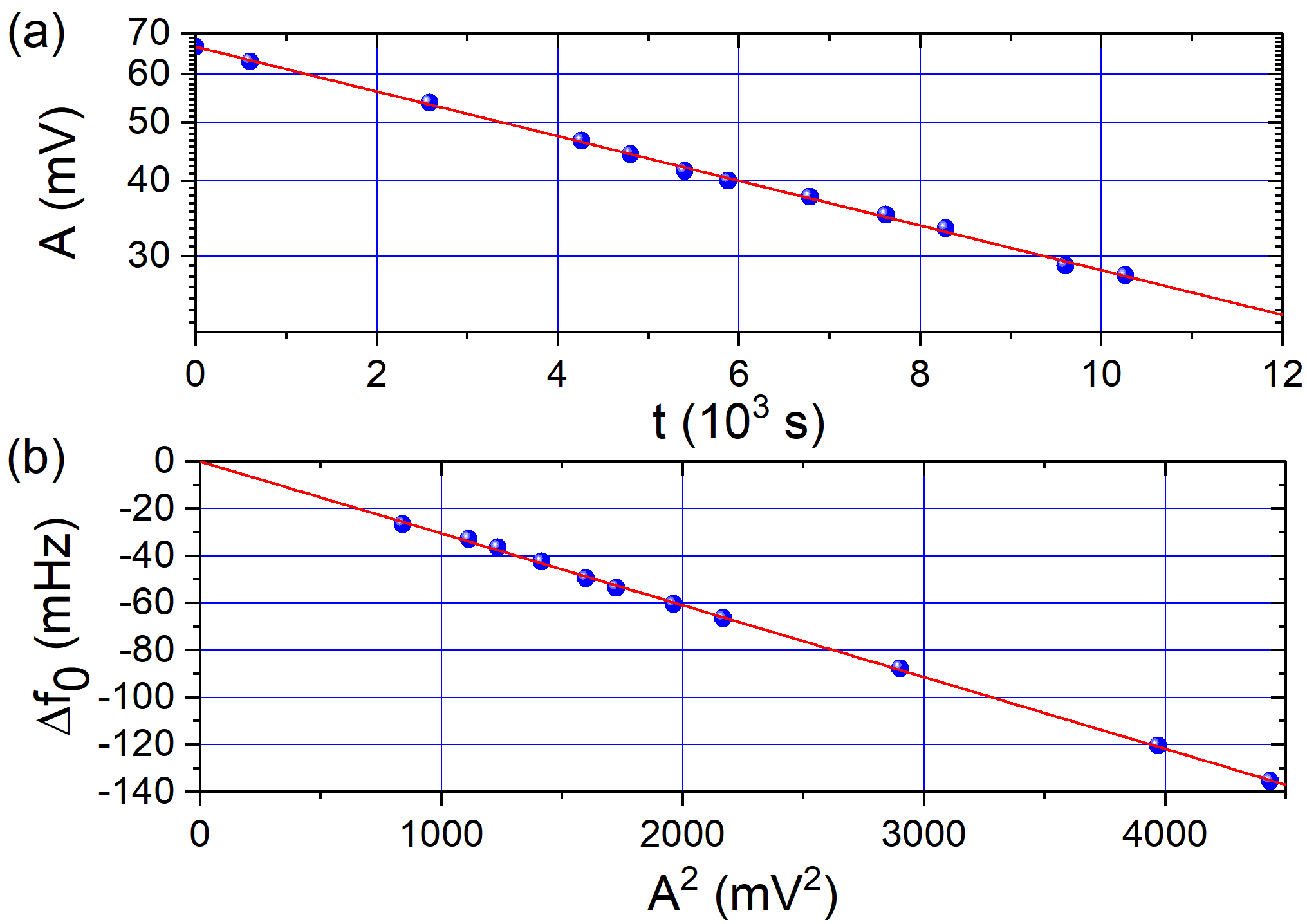}
\caption{(a) Ringdown measurement of the $z$ mode at the nominal pressure $P=1.0 \times 10^{-4}$ mbar. (b) Frequency shift of the $z$ mode as a function of the uncalibrated square amplitude, measured during a ringdown.}  \label{decay2}
\end{figure}

Let us discuss the behaviour of damping as a function of the gas pressure. For high pressure, $P>10^{-4}$~mbar, we find very similar $\tau$ for all $5$ modes, regardless of frequency. This is in agreement with standard gas damping models, which predict for translational and rotational (amplitude) damping in molecular regime the following expressions \cite{epstein,weber}:
\begin{align}  
 &\frac{1}{\tau_t} = \frac{1}{2} \left(1+\frac{8}{\pi} \right)\frac{P}{\rho R v_{\mathrm{th}}} \label{gasdamping1} \\
 &\frac{1}{\tau_r} = \frac{5}{\pi}\frac{P}{\rho R v_{\mathrm{th}}} 
 \label{gasdamping2}
\end{align}
where $P$ is the gas pressure, $v_{\mathrm{th}}=\sqrt{8 k_B T/ \pi m_g}$ the mean velocity of the gas molecules, with $m_g$ gas molecular mass, and $\rho$ and $R$ density and radius of the microsphere. The two expressions are indeed frequency-independent, depend linearly on pressure and differ one from each other by $\sim 10\%$. 

We have measured $\tau$ for the $z$ and $\beta$ modes at several pressures. The data is shown in Fig.~5. In order to compare the data with the gas damping model, we have to identify the pressure $P$ in Eqs.~(\ref{gasdamping1},\ref{gasdamping2}) with the pressure $P_c$ of the gas in the cold side of the vacuum chamber at $T_c=4.2$ K, where the microparticle is levitated. In general, $P_c$ differs from the nominal pressure $P_{w}$ measured by the pressure gauge at the warm side (room) temperature $T_{w}$, due to the so called thermomolecular effect \cite{sydoriak}. This phenomenon was studied in detail for the case of helium gas between $4$ K and $300$ K by Weber and Schmidt \cite{weberschmidt1,weberschmidt2} and interpolating formulae are reported in Ref.~\cite{sydoriak}. In Fig.~5 the pressure on the $x$-axis is the cold pressure $P_c$ obtained from the Weber-Schmidt model by specifying the measured warm pressure $P_w$ and the inner radius $r=0.97$ cm of the stainless steel tube which connects the cold chamber to the warm gauge pressure. In the limit of low pressure, the thermomolecular pressure ratio scales as $P_c/P_w=\left(T_c/T_w \right)^{\frac{1}{2}}$ and becomes independent of geometric factors.

The data for both the $\beta$ mode and the $z$ modes in Fig.~5 are fitted by a second-order polynomial fit significantly better than by a linear fit. We attribute the deviation from the expected linear behaviour to an imperfect estimation of $P_c$. The Weber-Schmidt formulae are obtained for ideal cylindrical connecting tubes, but in our experiment this is a crude approximation, due to the presence of the wiring inside the tube. However, as the low pressure limit of Weber-Schmidt formulae is independent of form factors of the connecting tube, we expect Eqs.~(\ref{gasdamping1},\ref{gasdamping2}) to be valid in this limit. Indeed, the linear terms obtained from the fits are respectively $\left( 5.6 \pm 0.3 \right)$ s$^{-1}$/mbar for the $\beta$ mode and $\left( 6.2 \pm 0.5 \right)$ s$^{-1}$/mbar for the $z$ mode. These estimations are in good agreement with the theoretical predictions $5.35$ s$^{-1}$/mbar and  $5.96$ s$^{-1}$/mbar obtained from Eqs.~(\ref{gasdamping1},\ref{gasdamping2}), using the measured values of $R$ and the nominal density of the material $\rho=7430$ kg/m$^3$ provided by the manufacturer. The agreement provides an indirect support to the Weber-Schmidt model. A notable feature of the gas damping model which is reproduced by the experiment is the ratio of translational to rotational damping of the order of $1.1$.

For the $\beta$ mode the intercept of the fit $1/\tau_{\beta0} =\left( 4.3 \pm 0.2 \right) \times 10^{-5}$ $s^{-1}$ corresponds to a quality factor $Q=2.7 \times 10^7$ at vanishing gas pressure. For the $z$ mode the intercept of the fit $1/\tau_{z0} =\left( -5 \pm 6 \right) \times 10^{-6}$ $s^{-1}$ is instead compatible with zero. This indicates that the residual damping time of the $z$ mode is longer than $1$ day, while the $Q$ factor is at least of the order of that of the $\beta$ mode at some $10^7$.

\begin{figure}[!ht]
\includegraphics[width=8.6cm]{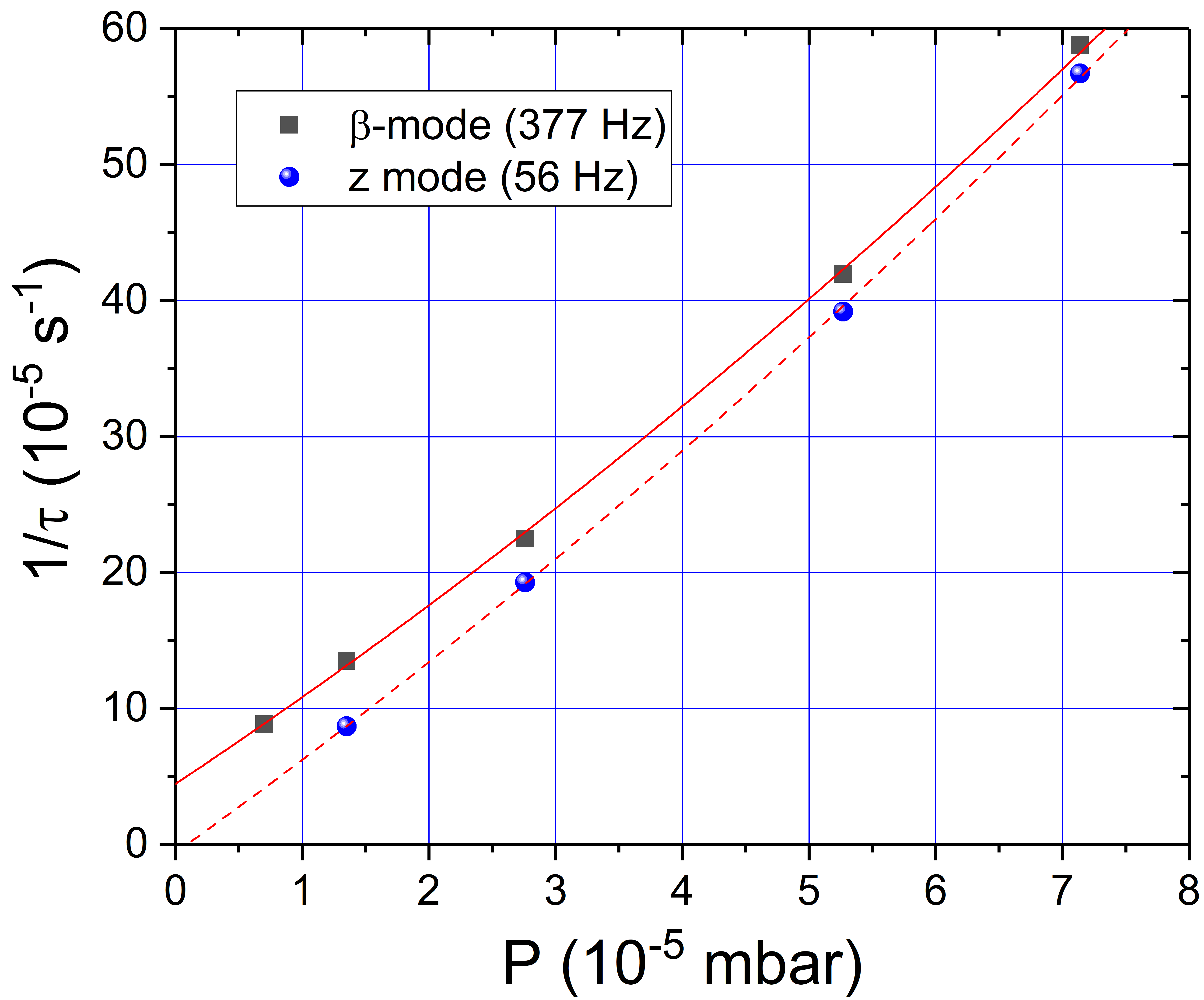}
\caption{Measured amplitude damping time as a function of the effective pressure at $T=4.2$~K calculated using the Weber-Schmidt model. The data refer to the $\beta$ mode (black squares) and $z$ mode (red circles). Solid and dashed lines are second order polynomial fits to the $\beta$ and $z$ mode data. }  \label{QvsP}
\end{figure}

\begin{figure}[!ht]
\includegraphics[width=8.6cm]{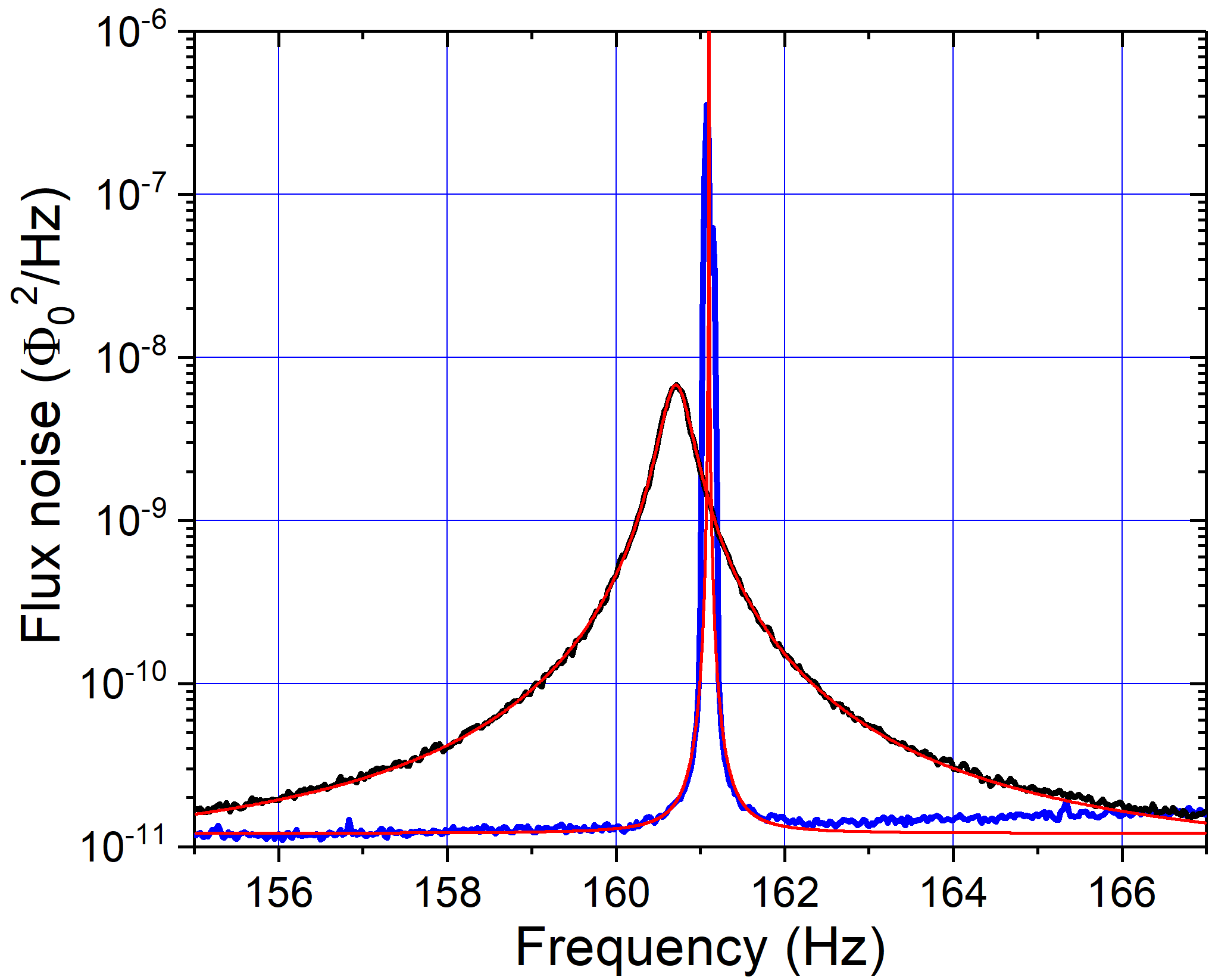}
\caption{Power spectrum of the flux measured by the SQUID around the frequency of the $\alpha$ mode, at two representative nominal pressures, $P=1.0 \times 10^{-4}$~mbar (narrow curve) and $P=5.0 \times 10^{-2}$~mbar (broad curve). Lorentzian fits are also shown.}  \label{160Hz}
\end{figure}

Finally, we have performed single mode noise measurements, focusing on the $\alpha$ mode, which corresponds to a horizontal twisting motion as in a torsion pendulum (Fig. \ref{scheme}b). Compared to other modes, the $\alpha$ one appears less sensitive to ambient vibrational noise. Fig.~\ref{160Hz} shows the same spectra of Fig.~\ref{spectrum}, zoomed in around the $\alpha$ mode. The high pressure data can be accurately fitted by a Lorentzian curve:
\begin{equation}
  S_{\Phi}=A_0+A_1 \frac{f_0^4}{\left(f^2-f_0^2 \right) ^2 + \left(f f_0/Q \right)^2}
\end{equation}
which gives $Q=421$ and $A_1=3.82 \times 10^{-14}$ $\Phi_0^2$/Hz.
The parameter $A_1$ is proportional to the torque noise acting on the microsphere. In the thermal noise limit, the latter is given by the expression:
\begin{equation}
S_\mathcal{T}=4 k_B T I \frac{\omega_0}{Q}   \label{torque}
\end{equation}
so we expect $A_1$ to drop at low pressure due to increasing $Q$. Indeed, the data at low pressure feature a much lower value of $A_1$. The curve superimposed on the low pressure data in Fig.~\ref{160Hz} is a fit restricted to the tails of the peak. In fact, we have removed $9$ bins around resonance which are affected by spectral leakage, due to the peak being narrower than the spectrum resolution. From the fit we get $A_1= \left( 1.6 \pm 0.1 \right) \times 10^{-16}$ $\Phi_0^2$/Hz. The excess noise above the resonance frequency originates from a broad peak produced in the SQUID electronics and causes a slight overestimation of the Lorentzian peak amplitude.

The fact that $A_1$ is much lower at low pressure than at high pressure is clearly indicating that the latter data are dominated by thermal noise from gas collisions. Based on this assumption, we estimate the torque noise at high pressure using Eq.~(\ref{torque}), $\sqrt{S_\mathcal{T}}=1.00 \times 10^{-20}$ Nm/$\sqrt{\mathrm{Hz}}$. This provides an absolute calibration factor to convert the $A_1$ factor measured at low pressure into a torque. From this we estimate the torque noise at low pressure as $\sqrt{S_\mathcal{T}}=\left( 6.4 \pm 0.2 \right) \times 10^{-22}$ Nm/$\sqrt{\mathrm{Hz}}$.

\section{Discussion}

The comparison between $\beta$ mode and $z$ mode suggests that when the pressure is reduced towards $0$ the dissipation moves from a gas damping regime, where the damping $1/\tau$ is independent of frequency, to a regime in which $1/\tau$ grows with frequency. This may correspond for instance to a loss angle $1/Q$ either constant or increasing with frequency. Under this regime the modes at low frequency should then achieve a $Q$ of the order of $10^7$ or higher. For a mode at $3-4$ Hz, as the $y$ mode in the present experiment, this would translate to a damping time $\tau \approx 10^6$ s. 

Besides the obvious damping related with the gas, which can be in principle reduced to negligible levels by operating at lower pressure and temperature, we still need to discuss the origin of the residual dissipation, which is apparent at least in the $\beta$ mode data.  We consider magnetic and eddy current losses in the ferromagnetic sphere and losses in the superconductor as possible options.

Estimations of the first two sources are reported in Appendix 3. For our particle, the order of magnitude of eddy current losses is $Q\approx 10^{11}$, so we can reasonably rule them out as the dominant effect. On the other hand magnetic hysteresis losses in the ferromagnetic particle could give the right order of magnitude, assuming a magnetic susceptibility with imaginary part of order $10^{-3}$. 

A hint on the limiting loss mechanism could be given by the dependence of the $Q$ factor on the microparticle size. We have performed additional measurements on a second similar particle with radius $a=39$ $\mu$m finding a residual $Q=6 \times 10^6$ for the $\beta$ mode. Other measurements performed with a different type of magnetic particle, such as cylinders with size larger of $200\times 400$ $\mu$m have shown $Q$ factors below $10^6$. The general trend is therefore towards an improvement of $Q$ by lowering the particle size. As shown in Appendix $3$, magnetic losses in the particle are in principle consistent with this picture.

Finally let us consider possible effects arising from nonidealities in the superconducting lead trap. Under ideal Meissner effect we do not expect significant losses from the superconductor, due to oscillation frequency being many orders of magnitude smaller than the superconducting gap frequency \cite{romero3}. However, it is notoriously difficult to observe ideal Meissner effect in real type-I superconducting samples \cite{meissner}. Flux penetration and mixed states can arise even in very pure lead polycrystalline samples. Normal metal regions arising from partial field penetration may in principle cause additional eddy current dissipation \cite{budker1}.

In order to investigate a possible role of flux penetration in our experiment, we have performed additional measurements by applying an external field to the lead trap during the superconducting transition. The field is applied through a superconducting coil placed outside the lead trap. Flux penetration appears as a shift of the $z$ and $\beta$ modes with respect to the typical values obtained in absence of applied field. We find that flux penetration shows up at fields down to $50$ $\mu$T, comparable with the Earth magnetic field and much lower than the nominal critical field of lead $H_c=80$ mT. The amount of flux penetration and frequency shift is not reproducible upon thermal cycling of the superconductor. For large frequency shifts, above $20 \%$,  we also observe a strong increase of dissipation, up to 2-3 orders of magnitude. This finding indicates that a strong attenuation of any external field is crucial to achieve high $Q$. In our experiment the magnetic shielding reduces the Earth field to the level of a few $\mu$T. This is one order of magnitude lower than the field necessary to cause significant flux penetration, indicating that residual losses due to flux penetration are likely negligible in the current setup. An interesting corollary is that our experimental technique can be used to probe deviations from the ideal Meissner effect in a macroscopic sample of a type I superconductor, and to assess the conditions under which ideal Meissner effect can effectively show up.

Let us compare our results with other micro/nanomechanical techniques and experiments. While higher quality factors have been reported in several systems, both in optically levitated nanoparticles \cite{fb2} and in clamped ultrastressed membranes \cite{schliesser}, values exceeding $10^7$ are not easily obtained. Compared to previous attempts of magnetically levitating a ferromagnetic particle above a superconductor \cite{budker1,chris} we obtain quality factors $3$ orders of magnitude higher. 

The damping time $\tau$ is comparable with recent results obtained with optical \cite{monteiro}, diamagnetic \cite{durso} and electrical \cite{pontin} levitation, with the advantage of lower temperature. In terms of the thermal noise factor, our magnetically levitated particles achieve $T/\tau \approx 10^{-4}$ K/s, which is already at the state of the art for micro/nanosystems. Comparable values have been achieved so far in ultracold cantilevers \cite{vinanteCSL2}. However, we stress that our approach has a substantial potential for improvement by fully suppressing gas damping and further cooling to millikelvin temperature, with values below $10^{-6}$ K/s which appear to be within reach. 

By assuming a thermal noise limited operation, the microparticle described in this work would feature a force noise spectral density $\sqrt{S_f} \approx 1$ aN/$\sqrt{\mathrm{Hz}}$ for the $z$ mode. This noise level has been achieved so far only with much smaller masses \cite{monteiro, force1, pontin}. Therefore, our micromagnet would be suitable to detect forces which scale with the size or with the mass of the particle. We can also consider the potential use as compact accelerometer, in particular as subresonant gravimeter \cite{goodkind}, despite the relatively low mass compared to macroscopic systems. For operation at $T=4.2$ K and with a lowest mode at $1$ Hz with $Q=3 \times 10^7$, we estimate a thermal noise limited acceleration noise of $3 \times 10^{-10}$ m/s$^2/\sqrt{\mathrm{Hz}}$, which would be competitive with state-of-the-art gravimeters \cite{goodkind, atomgrav}. Due to the compactness of our setup, an attractive possibility is the realization of gravity gradiometers composed of several pairs of identical trapped particles \cite{gradiometers}. 

In order to exploit this potential for sensing applications, some  technical issues need to be addressed. First, vibrational noise has to be substantially suppressed below the thermal noise. For resonant force detection in the $1-100$ Hz, seismic noise has to be suppressed by at least $2$ orders of magnitude. Second, feedback cooling of all normal modes will be necessary to shorten the measurement time and ensure stable operation in the linear regime. This can be done inductively by means of the driving coil, or electrostatically, by means of nearby electrodes and residual charge on the magnet. Finally, one may substantially reduce the displacement noise in order to increase the measurement bandwidth. This can be achieved by improving the SQUID sensor and optimizing of the pick-up coil coupling.

Beside direct sensing of forces or acceleration, we discuss now the unique potential of a levitated micromagnet as torque-based magnetometer. We limit our discussion to the $\alpha$ mode, and assume the magnetic moment $\bm{\mu}$ is oriented along the $y$ axis. By applying a magnetic field $\bm{B}$ along $x$, one would generate a torque $\mathcal{T}=\mu B$ directly driving the $\alpha$ mode. From the torque noise of the $\alpha$ mode estimated from data shown in Fig.~\ref{160Hz}, we infer an effective magnetic field noise $\sqrt{S_B}=\sqrt{S_\mathcal{T}}/\mu=14$ fT/$\sqrt{\mathrm{Hz}}$. This figure is comparable with that of SQUID \cite{SQUIDmagnetometers} or atomic \cite{atomicmagnetometers} magnetometers with much larger characteristic size of the order of $1$ mm. Therefore, our levitated particle has the potential to outperform existing state-of-the-art magnetometers in terms of field resolution normalized over the sensed magnetic field volume. This finding can be stated more precisely in terms of the so called quantum limit of magnetometry \cite{mitchell}, which can be defined through the relation:
\begin{equation}
  \frac{S_B}{2 \mu_0}= \frac{\hbar}{V}   \label{ql}
\end{equation}
where $V$ is the volume. Eq.~(\ref{ql}) expresses the fact that the magnetic field resolution scales inversely with the averaging volume. By taking the physical volume of our microsphere as reference, the quantum limited magnetic field resolution can be calculated for our experiment as $\sqrt{S_{B\mathrm{,QL}}}= 57$ fT/$\sqrt{\mathrm{Hz}}$. This means that, in contrast with other magnetometers \cite{mitchell}, our micromagnet could potentially beat the quantum limit by a factor of $4$. By achieving the theoretical thermal motion of the $\beta$ mode with the measured $Q$ factor, the field resolution would be $\sqrt{S_B}\approx 1$ fT/$\sqrt{\mathrm{Hz}}$ almost two orders of magnitude better than the quantum limit. A more detailed discussion of this unexpected finding goes beyond the scope of this paper. In particular, it is questionable whether the microsphere physical volume is the proper normalization factor to be considered in Eq.~(\ref{ql}). However, we point out that schemes employing levitated ferromagnets to overcome quantum limits on magnetometry have been theoretically proposed in literature \cite{budker2}. While the dc precession scheme proposed in \cite{budker2} is conceptually different from ours, the predicted potential performance is comparable, reaching some $2-3$ orders of magnitude below the quantum limit of Eq.~(\ref{ql}) under thermal noise limited operation.

From a practical point of view the magnetic microsphere cannot be used as a direct sensor of an external magnetic field, because in the Meissner limit the trap would completely shield the field. However, we can envision a practical magnetometer where the magnetic field is externally sensed by a superconducting pick-up coil circuit and applied to the microsphere by a nearby input coil placed inside the trap, as routinely done in SQUID magnetometers. In this case, some loss of efficiency due to geometrical couplings and partial image shielding will be unavoidable. However, the microsphere could be used as a direct sensor to probe pseudomagnetic fields such as those arising from exotic spin-dependent interactions of electrons \cite{budker2} or from interactions mediated by axion-like particles \cite{axions}. These interactions are not supposed to be shielded by the superconducting trap.


\section{Conclusions}
We have demonstrated that stable levitation of micromagnets using type-I superconducting traps, and measured ultralow damping of the rigid body mechanical modes. Due to the simultaneous combination of ultralow damping and low temperature, this system features very low thermal noise, opening the way to a number of extreme sensing applications, to detect weak forces, acceleration, torque and magnetic fields. For instance, this technology can be exploited in the search for exotic effects arising from new physics \cite{geraci, budker2, axions, monteiro}, measurements of the gravitational constant $G$ with small test masses \cite{aspelmeyer} and tests of collapse models \cite{vinanteCSL, vinanteCSL2, pontin}. In the context of quantum technologies, ultraisolated micromagnets can be strongly coupled to single spins \cite{hetet}, superconducting circuits \cite{romero1} or acoustomagnonic modes \cite{romero4}, enabling unconventional quantum protocols and, on more fundamental level, tests of quantum mechanics at mesoscopic scale \cite{romero2}.

Note: when finalizing this paper we became aware of a related paper by Gieseler et al. \cite{gieseler}.

\begin{acknowledgments}
We thank V. Sglavo, G. Guella, P. Connell and M. Utz for technical help, and M.W. Mitchell for discussions on magnetometry. We acknowledge financial support from the EU H2020 FET project TEQ (Grant No. 766900), the Leverhulme Trust (RPG-2016-046), the COST Action QTSpace (CA15220), INFN and the Foundational Questions Institute (FQXi).
All data supporting this study are openly available from the University of Southampton repository at https://doi.org/10.5258/SOTON/D1402. 
\end{acknowledgments}

\pagebreak

\appendix

\section{Numerical study of the System and Frequencies Characterization}

\begin{figure}
 \includegraphics[width=0.45\textwidth]{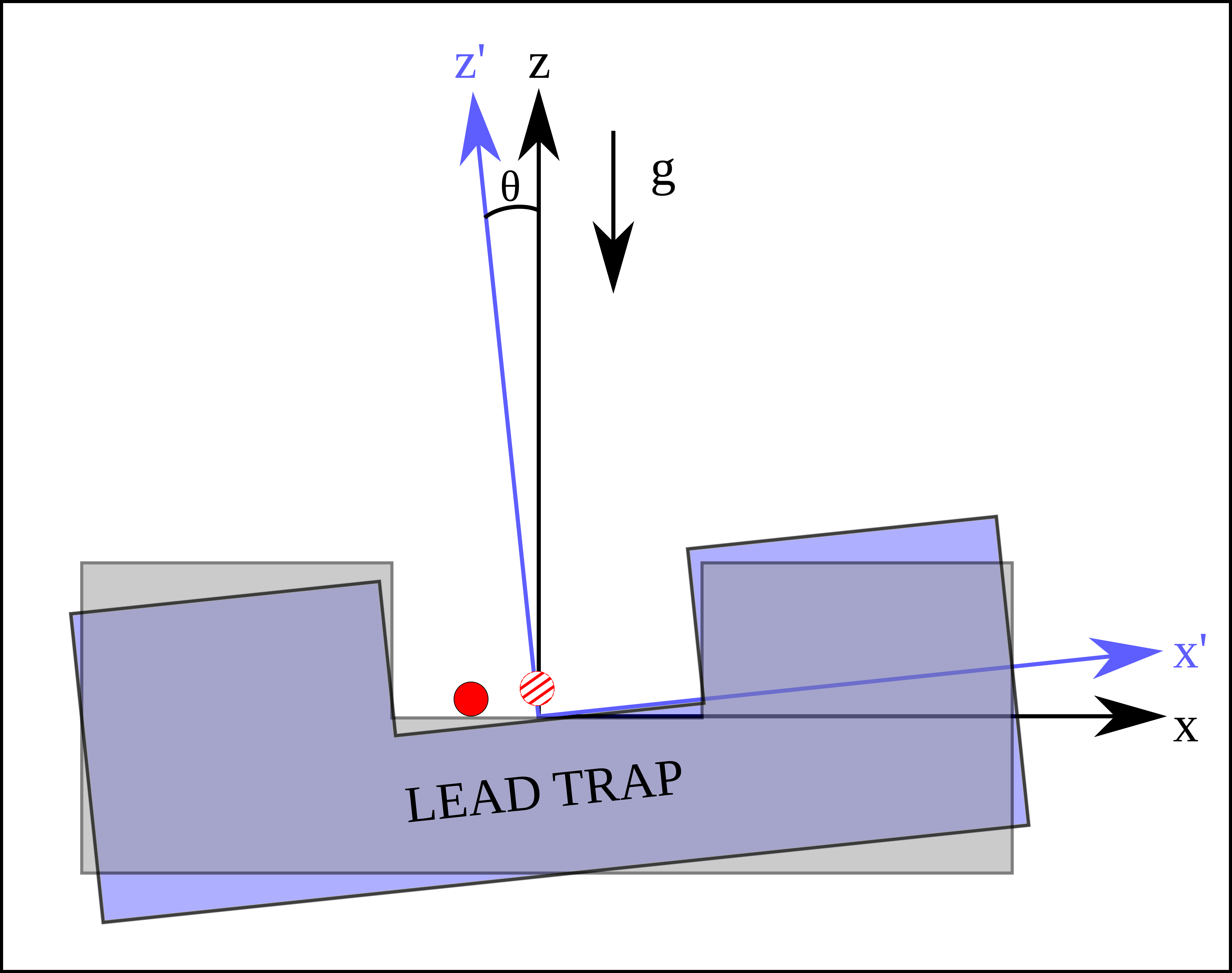}
  \caption{\label{Experiment} Simplified scheme of the trap with a tilt angle $\theta$.}{}
\end{figure} 
 
The potential energy of a ferromagnetic particle in the presence of a superconductor 
is described by
\begin{align}
\label{eq:poten}
U= -\frac{1}{2}\int_{V} d \bold{x} \bm{M}(\bold{x})\cdot \bold{B}_{\rm{ind}}(\bold{x})- m g z,
\end{align}
 where $\bm{M}$, $m$ and $V$ are the magnetization, the mass and the volume of the particle, and $\bold{B}_{ind}$ is the magnetic field induced by the presence of the superconductor.
Determining the induced magnetic field for a dynamical system is not an easy task, however as the motion of the magnet is much slower then the time scale of the super-currents and the London penetration depth is much smaller than the characteristic length scale of the system,  we can rely on a quasi-static approximation and assume the London penetration depth to be zero. Under this approximation the induced magnetic field $\bold{B}_{\rm{ind}}$ can be described by the following differential equation
\begin{eqnarray}\label{eq:max}
\left\{\begin{array}{cl}\bm{\nabla} \times \bm{B}_{\rm{ind}}& = 0\hspace{0.2cm} x \in \Omega  \nonumber\\
\bm{n}\cdot \bm{B}_{\rm{ind}}& =- \bm{n}\cdot \bm{B}_{\rm{f}}  \hspace{0.2cm} x\in \partial \Omega\end{array}\right.
\end{eqnarray} 
where $\partial \Omega$ represents the surface of the superconductor and $\Omega$ the space outside the superconductor, allowing us to numerically solve the magneto-static problem using a finite-element method with FEniCS software \cite{fenics}.

\begin{figure*}
  \begin{tabular}{cc}
    \includegraphics[width=0.48\textwidth]{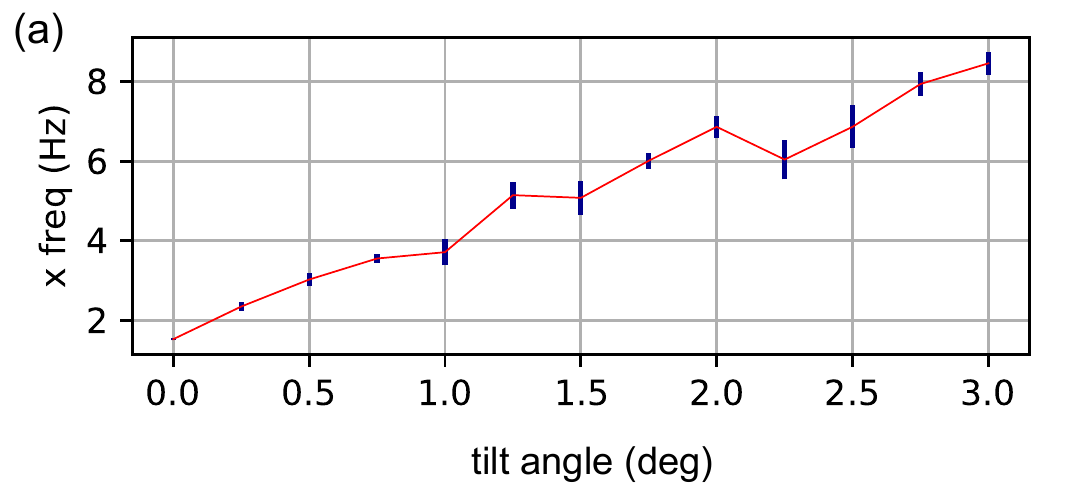} &   \includegraphics[width=0.48\textwidth]{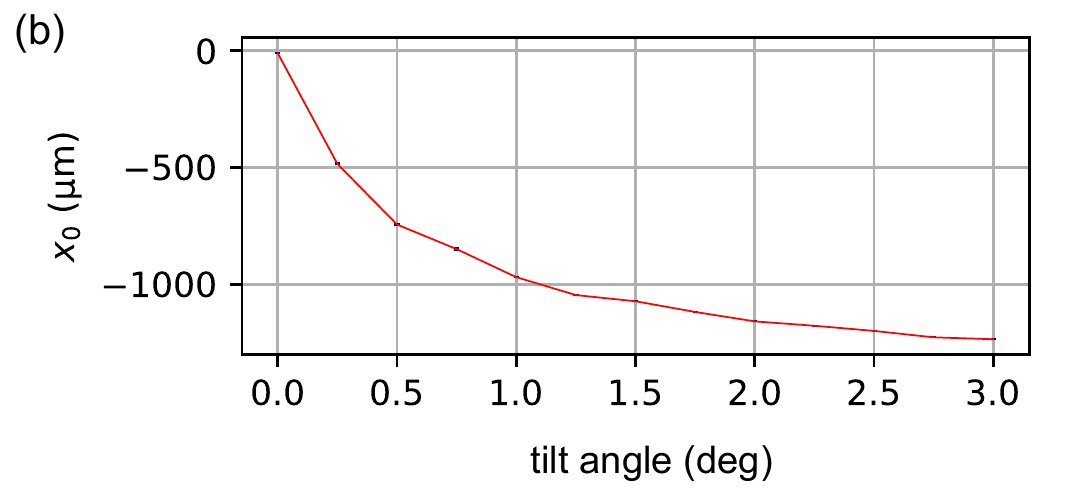}\\[4pt]
    \includegraphics[width=0.48\textwidth]{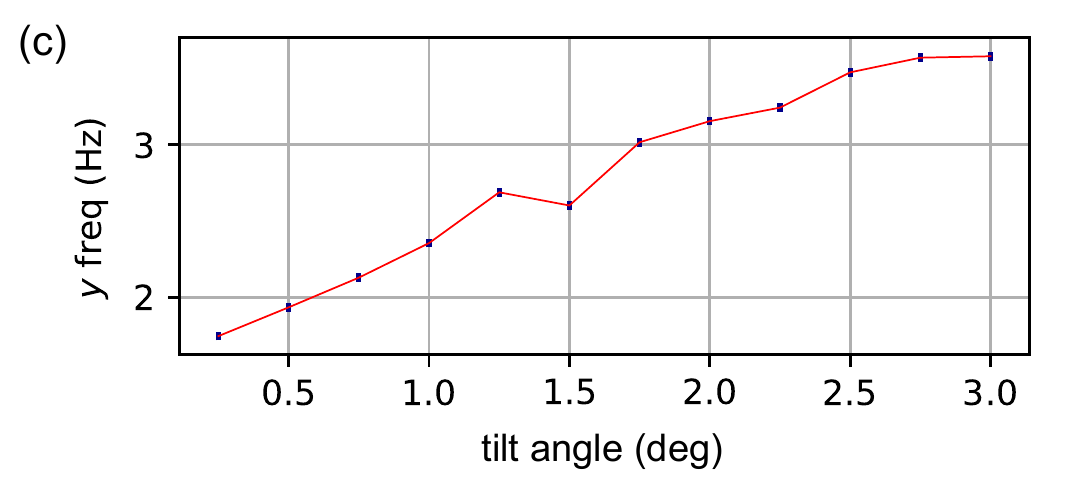} &   \includegraphics[width=0.48\textwidth]{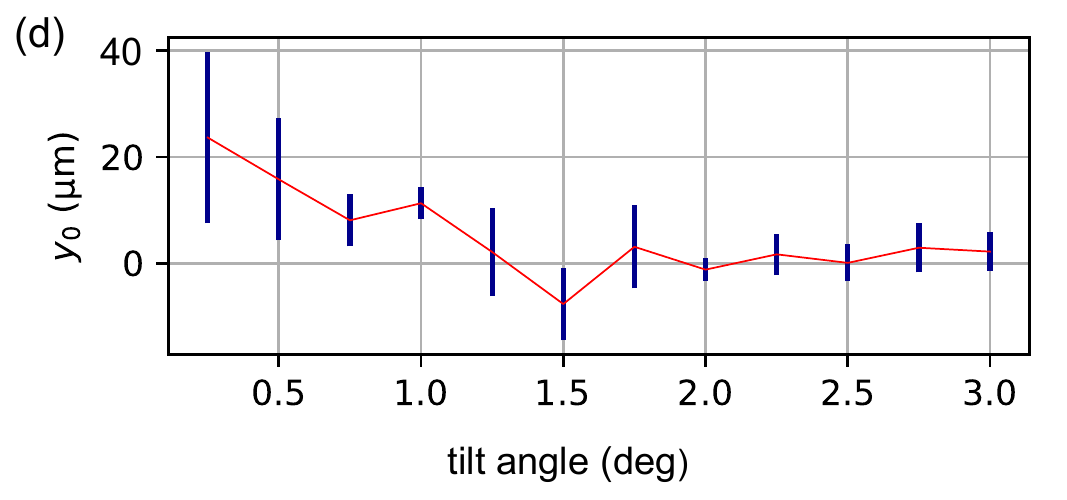} \\[4pt]
    \includegraphics[width=0.48\textwidth]{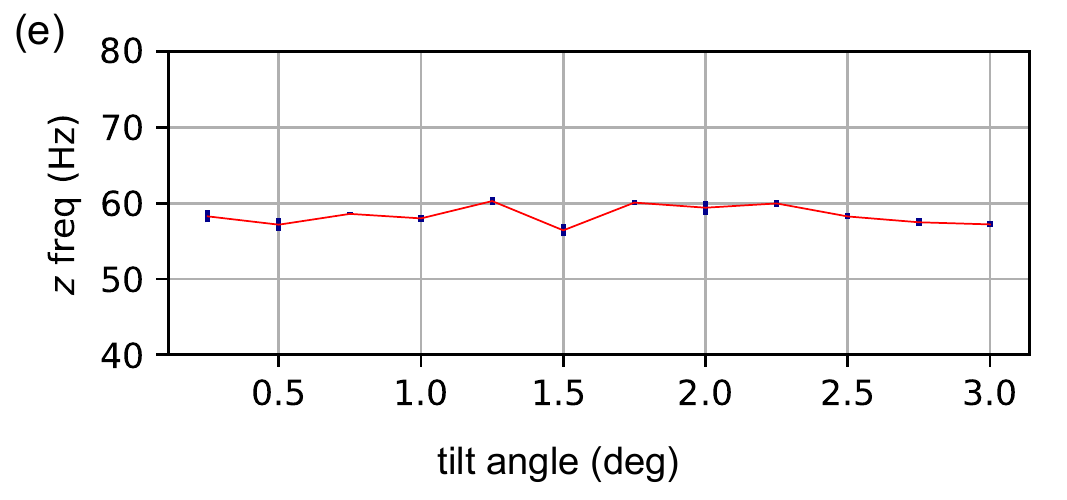} &   \includegraphics[width=0.48\textwidth]{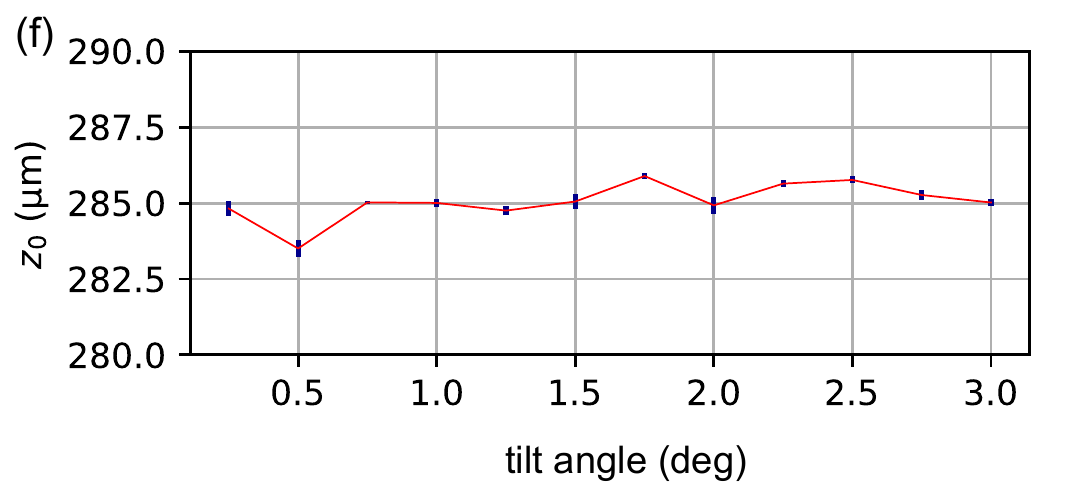} \\[4pt]
    \includegraphics[width=0.48\textwidth]{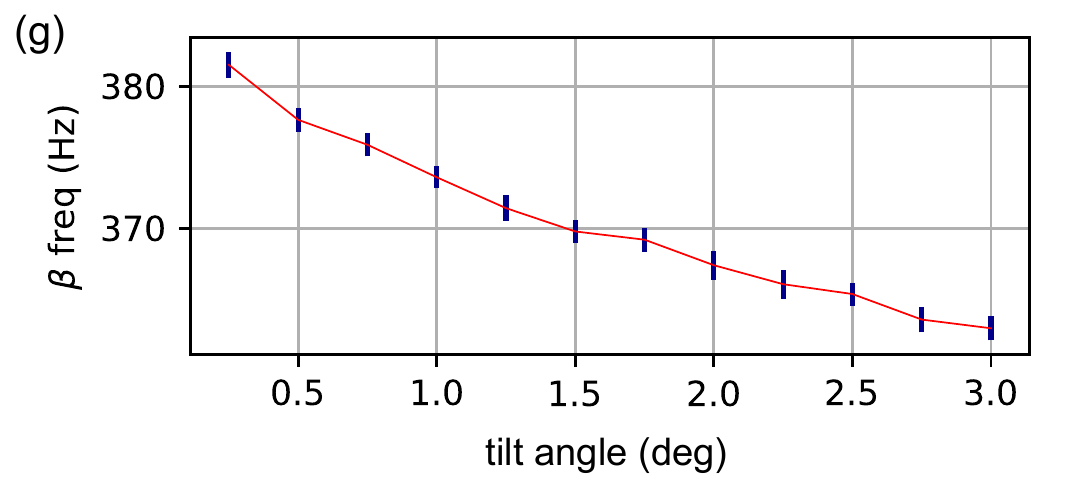} &   \includegraphics[width=0.48\textwidth]{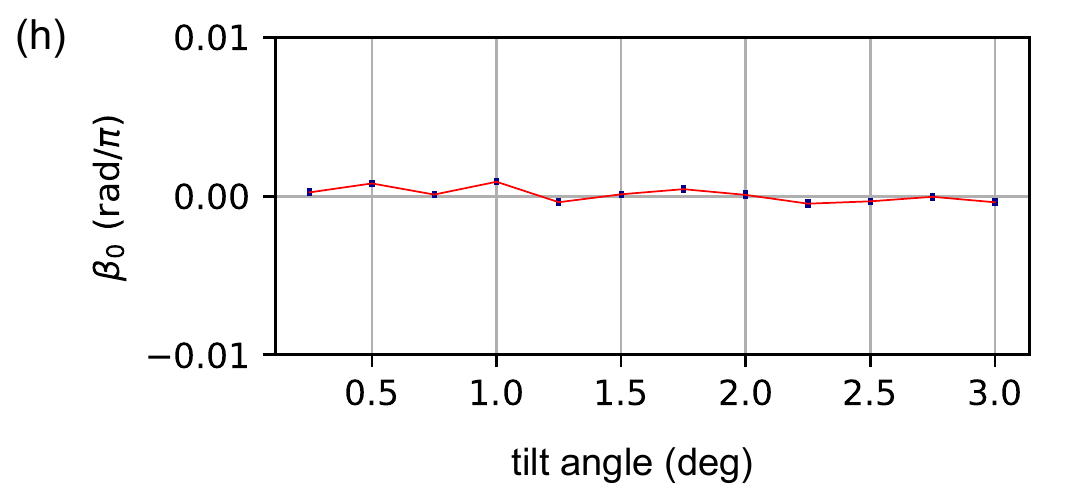}\\[4pt]
    \includegraphics[width=0.48\textwidth]{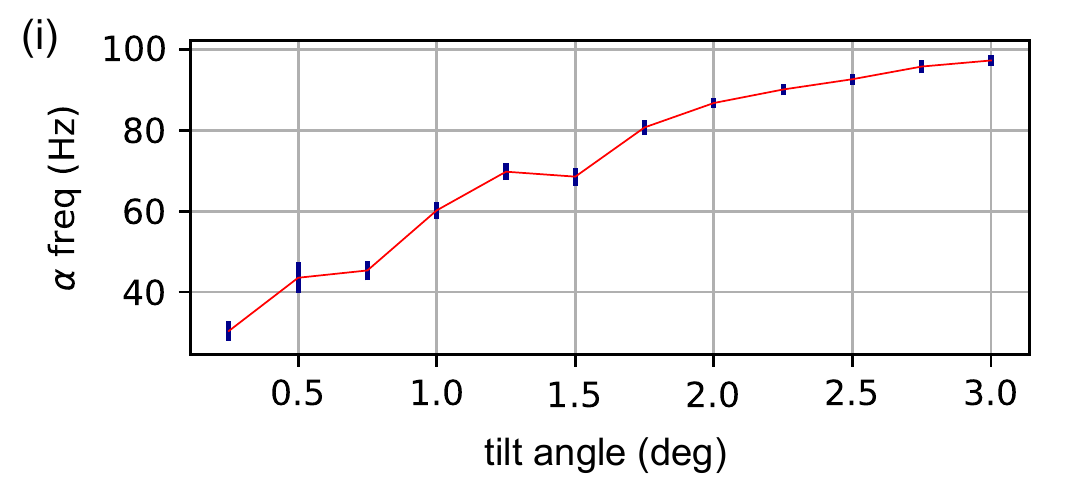} &   \includegraphics[width=0.48\textwidth]{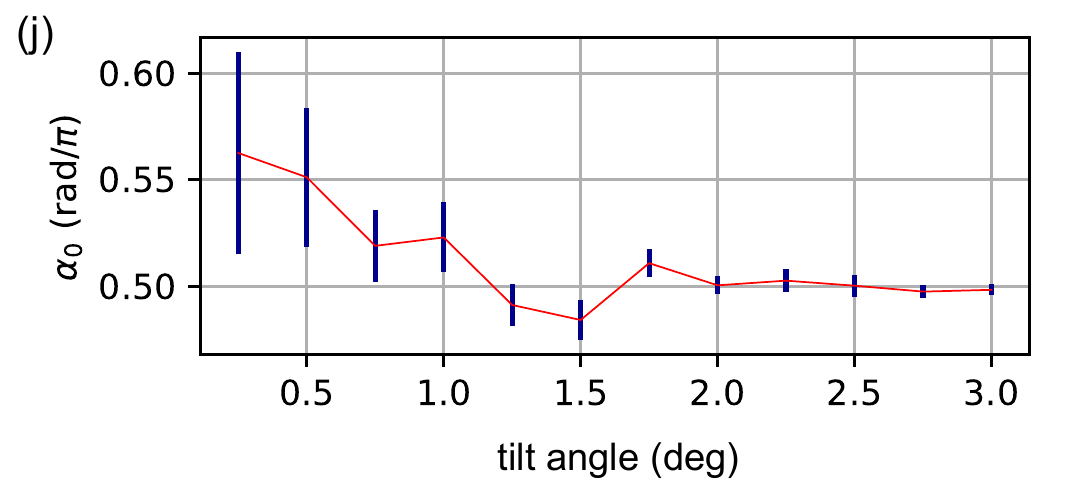}
  \end{tabular}
  \vspace{-0.2cm}
  \caption{\label{plots} Panels (a),(c),(e),(g),(i): dependence of the mode frequencies $(\omega_x, \omega_y, \omega_z, \omega_\beta, \omega_\alpha) /2\pi$ on the angle of tilt $\theta$ with respect to gravity. Panels (b),(d),(f),(h),(j): same for equilibrium positions $(x_0, y_0, z_0, \beta_0, \alpha_0)$.}
\end{figure*}

The simulation was carried out to find the equilibrium positions ($x_0, y_0, z_0, \beta_0, \alpha_0$) and the frequencies ($\omega_x, \omega_y, \omega_z, \omega_\beta, \omega_\alpha$) in the experimental setup with the $z$ axis tilted from $0^{o}$ to $3^{o}$ with respect to the gravitational axis, as shown in figure \ref{Experiment}. We extract the frequencies by fitting the data from the simulation around the minima with a quadratic function. Plots of these frequencies and minima are shown in figure \ref{plots}. 

\section{Calculation of the frequency shift}

The frequency shift as a function of the motion amplitude for an unforced mechanical resonator due to nonlinear restoring terms can be calculated using various ways, such as the Lindstedt-Poincare method or the multiple scales method, see for instance ref.~\cite{nayfeh} for an extensive review.

When the equation of motion is written in the following form, where dissipation terms are neglected and nonlinear terms up to the third order are retained:
\begin{equation}
  \ddot x + \omega_0^2 x +\alpha_2 x^2 + \alpha_3 x^3 =0
\end{equation}
the resulting effective resonance frequency is found to be:
\begin{equation}
  \tilde{\omega_0}=\omega_0 \left[ 1+\left( \frac{3}{8} \frac{\alpha_3}{\omega_0} -\frac{5}{12} \frac{\alpha_2^2}{\omega_0^4} \right) x_A^2 \right]   \label{freqshift}
\end{equation}
where $x_A$ is the amplitude of the oscillation, meaning that the oscillation is written as $x=x_A \mathrm{cos} \left( \omega_0 t +\phi_0 \right)$

For the $\beta$ and $z$ modes in the present experiment we can start from the potential predicted by the image method model, and  Taylor expand up to fourth order. We obtain the following equations of motion:
\begin{align}
 & \ddot z + \omega_z^2 z - \frac{5}{2}\frac{\omega_z^2}{z_0^2}z^2+5 \frac{\omega_z^2}{z_0^3} z^3 =0  \label{nonlinearzeta} \\
 & \ddot \beta + \omega_\beta^2 \beta -\frac{2}{3} \omega_\beta^2 \beta^3 =0 \label{nonlinearbeta}
\end{align}
where for each equation we have assumed that only that specific mode is excited at large amplitude, so that cross-coupling terms such as $\beta^2 z^2$ can be neglected.

Combining Eq.~({\ref{freqshift}}) with Eqs.~(\ref{nonlinearzeta},\ref{nonlinearbeta}) we finally obtain:
\begin{align}
 & \tilde{\omega_z}=\omega_z \left( 1 - \frac{35}{48 } \frac{z_A^2}{z_0^2} \right)  \\
 & \tilde{\omega_\beta}=\omega_\beta \left( 1 - \frac{1}{4} \beta_A^2  \right)   
\end{align}
where $z_A$ and $\beta_A$ are the oscillation amplitudes.
In both cases the nonlinear terms produce a softening effect proportional to the square of the amplitude of oscillation.
The frequency shift observed in the experiment can then be used to provide an approximate estimation of the absolute oscillation amplitude.

\section{Estimation of magnetic dissipation}

To estimate dissipation from eddy currents, we assume the microparticle has a finite electrical conductivity $\sigma$ and relative permeability $\mu_r\approx 1$. The latter condition is expected for a hard ferromagnetic particle fully saturated.
Under these assumptions the calculation of eddy current losses for a spherical geometry is a standard problem in electromagnetism \cite{landau}. Defining the complex polarizability $\alpha$ with the relation $\bm{ M} = \alpha \bm{ H}$, where $\bm{ M}$ is the magnetization induced by eddy currents and $\bm{ H}$ magnetic field, the induced magnetic moment is $\bm{ \mu_i}=\left( \alpha' + i \alpha'' \right)V \bm{ H} $. Following Landau-Lifshitz \cite{landau} in the low frequency limit $\delta \gg a$, where $\delta=\sqrt{2/\sigma \mu_0 \omega}$ is the penetration depth, condition largely fulfilled in our system:
\begin{equation}
  \alpha''=\frac{1}{5} \left( \frac{a}{\delta} \right)^2
\end{equation}
This leads to a power dissipation:
\begin{equation}
  W_{ec}=\frac{1}{2 \mu_0}\omega \alpha'' B^2 V=\frac{\pi}{15}\sigma \omega^2 a^5 B^2.  \label{powereddy}
\end{equation}
where $V=4\pi a^3 /3$ is the particle volume.
The same result can be derived by a direct integration of the power dissipation, under the condition of complete field penetration and dividing the sphere in elementary conductive rings. 

In order to calculate the mechanical energy dissipated by eddy currents in the $\beta$ mode, we note that a rotation of the magnetic particle by an angle $\beta$ from the equilibrium position is equivalent to a change of the image magnetic moment $\Delta \bm{ \mu'}= -\mu \beta \bm{\hat z}$, which produces a change of the $B$ field orthogonal to $\bm{ \mu}$:
\begin{equation}
  \Delta B=\frac{\mu_0}{4 \pi}\frac{\mu \beta}{\left( 2 z_0 \right) ^3}
\end{equation}   
For a sinusoidal oscillation of $\beta$ we can identify $\Delta B$ with $B$ in Eq.~(\ref{powereddy}), and taking into account that the mechanical energy of the mode is $E=\frac{1}{2}k_\beta \beta^2$ we can finally calculate the loss angle as:
\begin{equation}
\frac{1}{Q_{ec}}=\frac{W_{ec}}{\omega E}  \label{1Q}
\end{equation}
Inserting the parameters of our experiment and assuming a typical value $\sigma \approx 10^6$ $\Omega^{-1}$m$^{-1}$ for NdFeB alloys, we obtain $Q_{ec}\approx 10^{11}$. This is clearly too high to explain the measured residual loss.

Magnetic hysteresis losses can be estimated following the same scheme, but now replacing the eddy current polarizability with a complex magnetic susceptibility $\chi=\chi'+i\chi''$. As we are using a fully saturated hard ferromagnet, we expect $|\chi| \ll 1$. The total power dissipation will be given  by:
\begin{equation}
  W_{m}=\frac{1}{2 \mu_0}\omega \chi'' B^2 V.  \label{powermag}
\end{equation}
and we can calculate $1/Q_m$ using Eq.~(\ref{1Q}) with $W_m$ in place of $W_{ec}$. Carrying out the calculation we find:
\begin{equation}
 \frac{1}{Q_{m}}=\frac{1}{24} \chi'' \left( \frac{a}{z_0} \right) ^3  \label{powermag2}
\end{equation}
Assuming that $\chi''$ is frequency independent and taking into account that $z_0 \propto a^{\frac{3}{4}}$, this implies $1/Q_m \propto a^\frac{3}{4}$, which means that we expect a slightly sublinear increase of $Q$ when reducing the radius $a$.
However, if $\chi'' \propto \omega$ and taking into account that $\omega_\beta \propto a^{-\frac{5}{8}}$ we find a nearly size-independent behaviour $1/Q_m \propto a^\frac{1}{8}$.


\begin{thebibliography}{<99>}

\bibitem{optoreview} M. Aspelmeyer, T.J. Kippenberg, F. Marquardt, \textit{Cavity Optomechanics}, Rev. Mod. Phys. 86, 1391 (2014).

\bibitem{martinis} A.D. O'Connell, M. Hofheinz, M. Ansmann, Radoslaw C. Bialczak, M. Lenander, Erik Lucero, M. Neeley, D. Sank, H. Wang, M. Weides, J. Wenner, John M. Martinis, and A.N. Cleland, \textit{Quantum ground state and single-phonon control of a mechanical resonator}, Nature 464, 697 (2010).

\bibitem{bouwmeester} W. Marshall, C. Simon, R. Penrose, and D. Bouwmeester, \textit{Towards Quantum Superpositions of a Mirror}, Phys. Rev. Lett. 91, 130401 (2003).

\bibitem{romero2} H. Pino, J. Prat-Camps, K. Sinha, B. P. Venkatesh, and O. Romero-Isart, \textit{On-chip quantum interference of a superconducting microsphere}, Quant. Sci. Techn. 3, 025001(2018).


\bibitem{afm} G. Binnig, C.F. Quate, and Ch. Gerber, \textit{Atomic Force Microscope}, Phys. Rev. Lett. 56, 930 (1986).

\bibitem{rugarsingle} D. Rugar, R. Budakian, H.J. Mamin, and B.W. Chui, \textit{Single spin detection by magnetic resonance force microscopy}, Nature 430, 329 {2004}.

\bibitem{bachtold} J. Chaste, A. Eichler, J. Moser, G. Ceballos, R. Rurali, and A. Bachtold, \textit{A nanomechanical mass sensor with yoctogram resolution}, Nature Nanotechn. 7, 301 (2012).

\bibitem{pike} H. Liu, W.T. Pike, C. Charalambous, A.E. Stott, \textit{Passive Method for Reducing Temperature Sensitivity of a Microelectromechanical Seismic Accelerometer for Marsquake Monitoring Below 1 Nano-g}, Phys. Rev. Appl. 12, 064057 (2019).

\bibitem{rowan} R.P. Middlemiss, A. Samarelli, D.J. Paul, J. Hough, S. Rowan, and G.D. Hammond, \textit{Measurement of the Earth tides with a MEMS gravimeter}, Nature 531, 614 (2016).

\bibitem{tang} S. Tang, H. Liu, S. Yan, X. Xu, W. Wu, J. Fan, J. Liu, C. Hu, L. Tu, \textit{A high-sensitivity MEMS gravimeter with a large dynamic range}, Microsyst. and Nanoeng. 5, 45 (2019).

\bibitem{budker2} D.F. Jackson Kimball, A.O. Sushkov, and Dmitry Budker, \textit{Precessing Ferromagnetic Needle Magnetometer}, Phys. Rev. Lett. 116, 190801 (2016).

\bibitem{vinanteCSL} A. Vinante, M. Bahrami, A. Bassi, O. Usenko, G. Wijts, and T.H. Oosterkamp, \textit{Upper Bounds on Spontaneous Wave-Function Collapse Models Using Millikelvin-Cooled Nanocantilevers}, Phys. Rev. Lett. 116, 090402 (2016).


\bibitem{tau} Note that we use here the symbol $\tau$ for the amplitude damping time, which is the quantity physically measured in the experiment, rather than the energy damping time which is a factor 2 smaller.



\bibitem{schliesser} D. Mason, J. Chen, M. Rossi, Y. Tsaturyan, and A. Schliesser, \textit{Continuous force and displacement measurement below the standard quantum limit}, Nature Phys. 15, 745 (2019).

\bibitem{painter} G.S. MacCabe, H. Ren, J. Luo, J.D. Cohen, H. Zhou, A. Sipahigil, M. Mirhosseini, O. Painter, \textit{Phononic bandgap nano-acoustic cavity with ultralong phonon lifetime}, arXiv:1901.04129.

\bibitem{lehnert} J.D. Teufel, T. Donner, D. Li, J.W. Harlow, M.S. Allman, K. Cicak, A.J. Sirois, J.D. Whittaker, K.W. Lehnert, and  R.W. Simmonds, \textit{Sideband cooling of micromechanical motion to the quantum ground state}, Nature 475, 359 (2011). 

\bibitem{rugar} H.J. Mamin and D. Rugar, \textit{Sub-attonewton force detection at millikelvin temperatures}, 
Appl. Phys. Lett. 79, 3358 (2001).

\bibitem{vinanteCSL2} A. Vinante, R. Mezzena, P. Falferi, M. Carlesso, and A. Bassi, \textit{Improved Noninterferometric Test of Collapse Models Using Ultracold Cantilevers}, Phys. Rev. Lett. 119, 110401 (2017).

\bibitem{chang} D.E. Chang, C.A. Regal, S.B. Papp, D.J. Wilson, J. Yeb, O. Painter, H.J. Kimble, and P. Zoller, \textit{Cavity opto-mechanics using an optically levitated nanosphere}, PNAS 107, 1005
(2010).

\bibitem{ashkin} A. Ashkin, J.M. Dziedzic, J.E. Bjorkholm, and S. Chu, \textit{Observation of a single-beam gradient force optical trap for dielectric particles}, Opt. Letters 11, 288 (1986).





\bibitem{fb2} J. Gieseler, B. Deutsch, R.Quidant, and L. Novotny, \textit{Subkelvin Parametric Feedback Cooling of a Laser-Trapped Nanoparticle}, Phys. Rev. Lett. 109, 103603 (2012).




\bibitem{geraci} Gambhir Ranjit, Mark Cunningham, Kirsten Casey, and Andrew A. Geraci, \textit{Zeptonewton force sensing with nanospheres in an optical lattice}, Phys. Rev. A 93, 053801 (2016).

\bibitem{force1} D. Hempston, J. Vovrosh, M. Toro\^{s}, G. Winstone, M. Rashid, H. Ulbricht, \textit{Force sensing with an optically levitated charged nanoparticle}, Appl. Phys. Lett. 111, 133111 (2017).

\bibitem{tongli} Thai M. Hoang, Yue Ma, Jonghoon Ahn, Jaehoon Bang, F. Robicheaux, Zhang-Qi Yin, and Tongcang Li, \textit{Torsional Optomechanics of a Levitated Nonspherical Nanoparticle}, Phys. Rev. Lett. 117, 123604 (2016).

\bibitem{force2} M. Rashid, M. Toro\^{s}, A. Setter, H. Ulbricht, \textit{Precession Motion in Levitated Optomechanics}, Phys. Rev. Lett. 121, 253601 (2018).

\bibitem{monteiro} Fernando Monteiro, Wenqiang Li, Gadi Afek, Chang-ling Li, Michael Mossman, and David C. Moore, \textit{Force and acceleration sensing with optically levitated nanogram masses at microkelvin temperatures}, arXiv:2001.10931.

\bibitem{romero1} O. Romero-Isart, L. Clemente, C. Navau, A. Sanchez, and J.I. Cirac, \textit{Quantum Magnetomechanics with Levitating Superconducting Microspheres}, Phys. Rev. Lett. 109, 147205 (2012).

\bibitem{paul} W. Paul, \textit{Electromagnetic traps for charged and neutral particles}, Rev. Mod. Phys. 62, 531 (1990).

\bibitem{pontin} A. Pontin, N.P. Bullier, M. Toro\v{s}, P.F. Barker, \textit{An ultra-narrow line width levitated nano-oscillator for testing dissipative wavefunction collapse}, arXiv:1907.06046.

\bibitem{teq} A. Vinante, A. Pontin, M. Rashid, M. Toro\v{s}, P.F. Barker, H. Ulbricht, \textit{Testing collapse models with levitated nanoparticles: Detection challenge}, Phys. Rev. A 100, 012119(2019).

\bibitem{hetet} P. Huillery, T. Delord, L. Nicolas, M. Van Den Bossche, M. Perdriat, G. Hetet, \textit{Spin-mechanics with levitating ferromagnetic particles}, Phys. Rev. B 101, 134415 (2020).

\bibitem{durso} B.R. Slezak, C.W. Lewandowski, J-F. Hsu and B. D'Urso, \textit{Cooling the motion of a silica microsphere in a magneto-gravitational trap in ultra-high vacuum}, New J. Phys. 20, 063028, (2018).

\bibitem{cinesi} Di Zheng, Yingchun Leng, Xi Kong, Rui Li, Zizhe Wang, Xiaohui Luo, Jie Zhao, Chang-Kui Duan, Pu Huang, Jiangfeng Du, Matteo Carlesso, and Angelo Bassi, \textit{Room temperature test of wave-function collapse using a levitated microoscillator}, Phys. Rev. Research 2, 013057 (2020).

\bibitem{chris} C. Timberlake, G. Gasbarri, A. Vinante, A. Setter, H. Ulbricht, \textit{Acceleration sensing with magnetically levitated oscillators}, Appl. Phys. Lett. 115, 224101 (2019).

\bibitem{budker1} T. Wang, S. Lourette, S.R. O'Kelley, M. Kayci, Y.B. Band, D.F. Jackson Kimball, A.O. Sushkov, and D. Budker, \textit{Dynamics of a Ferromagnetic Particle Levitated over a Superconductor}, Phys. Rev. Applied 11, 044041 (2019).

\bibitem{romero3} J. Prat-Camps, C. Teo, C.C. Rusconi, W. Wieczorek, and O. Romero-Isart, \textit{Ultrasensitive Inertial and Force Sensors with Diamagnetically Levitated Magnets}, Phys. Rev. Appl. 8, 034002 (2017).

\bibitem{romero4} C. Gonzalez-Ballestero, J. Gieseler, O. Romero-Isart, \textit{Quantum Acoustomechanics with a Micromagnet}, Phys. Rev. Lett. 124, 093602 (2020).

\bibitem{mitchell} M.W. Mitchell, S. Palacios Alvarez, \textit{Quantum limits to the energy resolution of magnetic field sensors}, Rev. Mod. Phys. 92, 21001 (2020).

\bibitem{magnequench} Magnequench, MQP-S-11-9-20001 isotropic powder, URL https://mqitechnology.com/

\bibitem{image1} J.D. Jackson, \textit{Classical Electrodynamics}, 3rd ed. Wiley, New York, 1998.

\bibitem{image2} Q.G. Lin, \textit{Theoretical development of the image method for a general magnetic source in the presence
of a superconducting sphere or a long superconducting cylinder}, Phys. Rev. B 74, 024510 (2006).

\bibitem{fenics} M.S. Aln{\ae}s, J. Blechta, J. Hake, A. Johansson, B. Kehlet, A. Logg, C. Richardson, J. Ring, M.E. Rognes, and G.N. Wells, \textit{The FEniCS Project Version 1.5}, Arch. Numer. Software 3, 9 (2015).

\bibitem{epstein} P.S. Epstein, \textit{On resistance experienced by spheres in their motion through gases}, Phys. Rev. 23, 710 (1924).

\bibitem{weber} A. Cavalleri, G. Ciani, R. Dolesi, M. Hueller, D. Nicolodi, D. Tombolato, S. Vitale, P.J. Wass, W.J. Weber, \textit{Gas damping force noise on a macroscopic test body in an infinite gas reservoir}, Phys. Lett. A 374, 3365 (2010).

\bibitem{sydoriak} T.R. Roberts and S.G. Sydoriak, \textit{Thermomolecular Pressure Ratios for He$^3$ and He$^4$}, Phys. Rev. 102, 304 (1956).

\bibitem{weberschmidt1} S. Weber, Leiden Comm. No. 264b (1936); 264d (1936); and Suppl. 71b (1932).

\bibitem{weberschmidt2} S. Weber and G. Schmidt, Leiden Comm. No  246c (1936).

\bibitem{meissner} W. Treimer, O. Ebrahimi, and N. Karakas, \textit{Observation of partial Meissner effect and flux pinning in superconducting lead containing non-superconducting parts}, Appl. Phys. Lett. 101, 162603 (2012). 

\bibitem{goodkind} John. M. Goodkind, \textit{The superconducting gravimeter}, Rev. Sci. Instrum. 70, 4131 (1999).


\bibitem{atomgrav} Zhong-Kun Hu, Bu-Liang Sun, Xiao-Chun Duan, Min-Kang Zhou, Le-Le Chen, Su Zhan, Qiao-Zhen Zhang, and Jun Luo, \textit{Demonstration of an ultrahigh-sensitivity atom-interferometry absolute gravimeter}, Phys. Rev. A 88, 043610 (2013).

\bibitem{gradiometers} C.E. Griggs, M.V. Moody, R.S. Norton, H.J. Paik, and K. Venkateswara, \textit{Sensitive Superconducting Gravity Gradiometer Constructed with Levitated Test Masses}, Phys. Rev. Appl. 8, 064024 (2017).


\bibitem{SQUIDmagnetometers} M. Schmelz, R. Stolz, V. Zakosarenko, T. Sch\"{o}nau, S. Anders, L. Fritzsch, M. M\"{u}ck, M. Meyer, H.-G. Meyer, \textit{Sub-fT/Hz$^{1/2}$ resolution and field-stable SQUID magnetometer based on low parasitic capacitance sub-micrometer cross-type Josephson tunnel junctions}, Physica C: Superconductivity and its Applications, 482, 27 (2012).

\bibitem{atomicmagnetometers} H.B. Dang, A.C. Maloof, and M.V. Romalis, \textit{Ultrahigh sensitivity magnetic field and magnetization measurements with an atomic magnetometer}, Appl. Phys. Lett. 97, 151110 (2010).

\bibitem{axions} N. Crescini, C. Braggio, G. Carugno, P. Falferi, A. Ortolan, G. Ruoso, \textit{The QUAX-g$_p$g$_s$ experiment to search for monopole-dipole Axion interaction}, Nucl. Instrum. Meth. Phys. Res. A 842, 109 (2017).


\bibitem{aspelmeyer} J. Schm\"{o}le, M. Dragosits, H. Hepach and M. Aspelmeyer, \textit{A micromechanical proof-of-principle experiment for measuring the gravitational force of milligram masses}, Class. Quantum Grav. 33, 12 (2016).



\bibitem{gieseler} J. Gieseler, A. Kabcenell, E. Rosenfeld, J.D. Schaefer, A. Safira, M.J.A. Schuetz, C. Gonzalez-Ballestero, C.C. Rusconi, O. Romero-Isart, M.D. Lukin, \textit{Single-Spin Magnetomechanics with Levitated Micromagnets}, Phys. Rev. Lett. 124, 163604 (2020).

\bibitem{nayfeh} A.H. Nayfeh and D.T. Mook, \textit{Nonlinear Oscillations}, Wiley Classic Library Classic Edition Published, 1995, pag. 50.

\bibitem{landau} L.D. Landau, E.M. Lifshitz, \textit{Electrodynamics of continuous media}, 2nd Edition, Pergamon Press, Oxford, pag. 205.



\end{thebibliography}
\end{document}